# Equiatomic quaternary Heusler alloys: a material perspective for spintronic applications

Lakhan Bainsla[#, *] and K. G. Suresh[#]

*Department of Physics, Indian Institute of Technology Bombay, Mumbai 400076, India*

**Abstract**

Half-metallic ferromagnetic (HMF) materials show high spin polarization and are therefore interesting to researchers due to their possible applications in spintronic devices. In these materials, while one spin sub band has a finite density of states at the Fermi level, the other sub band has a gap. Because of their high Curie temperature ($T_C$) and tunable electronic structure, HMF Heusler alloys have a special importance among the HMF materials. Full Heusler alloys with the stoichiometric composition $X_2YZ$ (where $X$ and $Y$ are the transition metals and $Z$ is a *sp* element) have the cubic $L2_1$ structure with four interpenetrating fcc sublattices. When each of these four fcc sublattices is occupied by different atoms ($XX'YZ$), a quaternary Heusler structure with different structural symmetry (space group *F*-43*m,* #216) is obtained. Recently, these equiatomic quaternary Heusler alloys (EQHAs) with 1:1:1:1 stoichiometry have attracted a lot of attention due to their superior magnetic and transport properties. A special class of HMF materials identified recently is known as spin gapless semiconductors (SGS). The difference in this case, compared to HMFs, is that the density of states for one spin band is just zero at the Fermi level, while the other has a gap as in the case of HMFs. Some of the reported SGS materials belong to EQHAs family. This review is dedicated to almost all reported materials belonging to EQHAs family. The electronic structure and hence the physical properties of Heusler alloys strongly depend on the degree of structural order and distribution of the atoms in the crystal lattice. A variety of experimental techniques has been used to probe the structural parameters and degree of order in these alloys. Their magnetic properties have been investigated using the conventional methods, while the spin polarization has been probed by point contact Andreev reflection (PCAR) technique. The experimentally obtained values of saturation magnetization are found to be in agreement with those estimated using the Slater-Pauling rule in most of the cases. Electrical resistivity and Hall measurements are being used to distinguish between SGS and HMF nature in detail. The current spin polarization value, $P = 0.70 \pm 0.01$ for CoFeMnGe is found to be highest among the EQHAs. CoFeMnSi and CoFeCrGa are found to show SGS behavior with high Curie temperatures, thus making them suitable substitutes for diluted magnetic semiconductors. CoRuFeSi is found to have the highest $T_C$ among EQHAs. Theoretical prediction of magnetic properties on the basis of electronic structure calculations has also been reported in a few systems, which are also discussed in this review. Thus, this review presents a consolidated picture of the magnetic and spintronic properties of this important, but relatively new class of Heusler alloys. It is expected that this will stimulate further interest in these alloys, thereby paving the way for the identification of more HMF and SGS materials. As a result of this, it is expected that more efficient spintronic devices using these alloys would emerge in the near future.

#Authors to whom correspondence should be addressed. Electronic address: suresh@phy.iitb.ac.in; lakhanbainsla@gmail.com

*Present address: WPI Advanced Institute for Materials Research, Tohoku University, Sendai 980-8577, Japan

**TABLE OF CONTENTS**

## I. INTRODUCTION

Spintronics has emerged as one of the most exciting topics of research in the field of magnetic materials in the recent past. The interest in this topic is two-fold; applied and fundamental. Focus on both these aspects has contributed significantly as revealed by the number of publications in the last five years. Novel methodologies are being adopted in the manipulation of spin degree of freedom of the electrons, which has led to a large number of novel and potential devices. The most important class of materials useful in spintronics is the half-metallic ferromagnets. The electronic band structure of these materials is different from that of typical metals (Fig. 1a) and semiconductors (Fig. 1b). Half-metallic ferromagnetic (HMF) materials possess a conducting channel for one spin and a semiconducting channel for the other, as shown in Fig. 1(c). HMF materials with 100% spin polarization are the ideal candidates for metallic spintronic applications, as they can give rise highly spin polarized currents [1-3]. HMF nature was predicted in several types of materials such as magnetic oxides [4], dilute magnetic III-V compound semiconductors [5], and Heusler alloys [2,3,6]. Among these, Heusler alloys have a special importance due to their higher Curie temperatures ($T_C$) and tunable electronic structure [2,3,6]. Since, the prediction of HMF nature in NiMnSb [6], Heusler alloys have received a special attention due to the reasons mentioned above. In the large family of Heusler alloys, generally Co based alloys show relatively high Curie temperature and high spin polarization [7-16].

HMF materials have potential applications as spin polarized current sources for current perpendicular to plane giant magnetoresistance (CPP-GMR) [17-19], magnetic tunnel junctions (MTJ) [20,21], non-local spin-valve (NLSV) devices [22,23] and spin injectors to semiconductors [24]. High magnetoresistance (MR) ratios and large spin accumulation signals in spintronic devices using Co-based Heusler alloys clearly reflect the half-metallic nature of these alloys [17-24]. Recently, a very high value of spin polarization was observed for Co-based alloys at low temperatures, but a strong degradation in the transport properties was seen at room temperature [15,17,19,20]. Therefore, further exploration of HMF Heusler alloys is strongly required both from application as well as fundamental points of view.

Recently, a new class of materials in the HMF family, namely spin gapless semiconductors (SGS) has started emerging and their role is expected to be vital in semiconductor spintronic devices [25-27]. SGS materials possess a zero band gap for one spin channel, while a band gap is present for the other channel, as shown in the Fig. 1(d).

Therefore, these materials can serve as a bridge between the HMFs and semiconductors. Some of the unique properties of SGS are: (i) spin polarized current resulting from electrons as well as spin polarized holes; (ii) ability to switch between *n* and *p* type spin polarized carriers by the application of an electric field; (iii) almost no energy required to excite electrons from the valence band to the conduction band. Due to their unique band structure, SGS materials are considered as possible candidates to substitute for diluted magnetic semiconductors (DMS). SGS materials have been receiving intense research interest since their theoretical prediction [25]. The major drawback of low Curie temperature in DMS [28,29] can be overcome by SGS materials since relatively high $T_C$ has been obtained for some SGS Heusler alloys.

Many Heusler alloys are found to possess the SGS band structure as revealed by the *ab-initio* calculations [27,30-32]. But, such a behavior is confirmed experimentally only in a few of them. i.e., $Mn_2CoAl$ [26], CoFeMnSi [8] and CoFeCrGa [10]. Very recently, thin films of $Mn_2CoAl$ were synthesized to check its applicability in the devices [33,34].

Full Heusler alloys (FHA), with the stoichiometric composition $X_2YZ$, where *X* and *Y* are the transition metals and *Z* is a *sp* (or main group) element, have the cubic $L2_1$ structure (space group *Fm-3m*, #225) with four interpenetrating fcc sublattices. When each of these sublattices is occupied by different atoms (*XX'YZ*), a quaternary Heusler structure with a different structural symmetry (space group *F-43m,* #216) is obtained. The resulting structure is named as LiMgPdSn type (or the so called Y-type structure as per Strukturberichte database). Depending on the occupation of the different lattice sites, three different types of structural configurations are possible for the Y-type structure [31,35]. A simple method to conceive a quaternary Heusler alloy such as ABCD with 1:1:1:1 stoichiometry is to consider it as the combination of two ternary full Heusler alloys $A_2BD$ and $A_2CD$. Because of the 1:1:1:1 stoichiometry, the quaternary alloys are termed as equiatomic quaternary Heusler alloys (EQHAs). From the application point of view, EQHAs have some advantages over pseudo-ternary alloys such as $X_2Y_{1-a}Y'_aZ$, since the random distribution of Y and Y' leads to an additional disorder scattering and thus the spin diffusion length becomes very short [36]. Since such a scattering is absent in EQHAs, the devices based on the EQHAs are expected to have low power dissipations.

In the following, a brief review of the structural, magnetic, magneto-transport, spin polarization and electronic structure properties of EQHAs is presented.

## II. STRUCTURAL PROPERTIES

Physical properties of Heusler alloys are strongly dependent on the atomic arrangements in the crystal. A slight disorder in the structure can alter the electronic structure distinctly. If each of the elements reside at their respective sites, the resultant will be a well-ordered cubic structure. Several types of disordered structures have been observed in Heusler alloys. Some of the possible disordered structures in FHA and/or EQHA are termed as A2, DO$_3$ and B2 structures [37]. The complete disorder in FHA and/or EQHA structure (X, X', Y and Z distributed randomly) results in the A2 structure with reduced symmetry and bcc lattice. On the other hand, the random distribution of X and Y or X and Z leads to the DO$_3$ disorder which results in a BiF$_3$-like structure. B2 type is another frequently observed structure, in which Y and Z sites become equivalent. i. e., the disorder is between Y and Z sites only, which leads to a CsCl-like structure. The crystal structures corresponding to various types of structures are given in Fig. 2. Table 1.1 summarizes the various possible disorders in the case of FHA and/or EQHA. Generally, it has been observed that a highly ordered crystal structure is essential to have a half-metallic electronic structure [2,37,38].

***Structural disorder and its determination:*** As mentioned above, the electronic structure and hence the physical properties strongly depend on the structural order and distribution of the atoms in the crystal lattice. Thus a careful analysis of the crystal structure is required to understand or even predict the properties of these materials. The band structure calculations show that even a small amount of atomic disorder can cause changes in their electronic structure and their magnetic and transport properties [38-40].

X-ray diffraction (XRD) technique is the easiest experimental method to examine the crystal structure. In the case of an ordered $L2_1$ phase, superlattice reflections (SR) (111) and (200) are usually present in the XRD data. A rough estimate of chemical order present in the materials can be obtained by calculating the $I_{200}/I_{220}$ and $I_{111}/I_{220}$ intensity ratios. The intensity of the superlattice reflections (111) and (200) are proportional to the order parameters, $S^2$ and $S^2(1-2\alpha)^2$, where $S = ((I_{200}/I_{220})_{Exp}/(I_{200}/I_{220})_{Theory})^{1/2}$ and $S(1-2\alpha) = ((I_{111}/I_{220})_{Exp}/(I_{111}/I_{220})_{Theory})^{1/2}$ [41]. In the case of A2 disorder, S = 0 and α = 0, for B2 disorder, S = 1 and α = 0.5. For a well ordered cubic phase, S = 1 and α = 0 are expected.

When Z element is from the same period as the transition metals, it is very difficult to find out the correct structure unambiguously by using x-ray or neutron diffraction data [42].

In such cases, XRD investigations with synchrotron radiation lead to better structural results, which allow the direct observation of the anti-site disorder. Mössbauer spectroscopy (MS) is another technique which can be used for structural investigations, as it measures the hyperfine field ($H_{hf}$) at the core Mössbauer active atoms such as Fe [43]. The nature of the obtained spectra indicates the local environment of the probed atomic species. Extended x-ray absorption fine structure (EXAFS) analysis is another technique, which has been successfully applied to investigate Heusler materials. It determines the short range chemical order around the atoms. For example, $Co_2MnSi$ is predicted to be HMF from band structure calculations in the ordered state; however, EXAFS analysis revealed a distinct amount of disorder between the Co and the Mn site, which can be the reason for the low measured spin polarization [44].

**A. CoFeMnZ (Z = Al, Ga, Si and Ge) alloys**

CoFeMnZ (Z = Al, Ga, Si and Ge) alloys are found to exist in the Y-type structure [45]. The room temperature powder XRD patterns of CoFeMnZ (Z = Si and Ge) are given in Fig. 3(*a*)-3(*b*). Alijani et al. [45] found some amount of structural disorder for CoFeMnAl and CoFeMnSi from x-ray diffraction measurements, while it is not confirmed for CoFeMnGe and CoFeCrGa due to similar scattering amplitudes of the constituent elements. To further investigate the crystal structure of CoFeMnZ (Z = Si and Ge) alloys, Bainsla *et* al. [7,8,46] performed EXAFS measurements [46] and found large structural disorder in the case of CoFeMnSi, while CoFeMnGe was found to exist in well ordered Y-phase. Room temperature $^{57}$Fe Mössbauer analysis also revealed a large structural disorder ($DO_3$) for CoFeMnSi [8]. Some amount of disorder was also obtained for CoFeMnGe form the room temperature Mössbauer analysis [7], which was contradictory to the EXAFS measurements [46]. Exact quantification of the disorder using the Mössbauer or the EXAFS technique is not possible because of the inherent inaccuracies of the fitting procedures used in these techniques. The structural stability of the CoFeMnZ (Z = Si and Ge) alloys was checked by Bainsla *et* al. [7,8], by performing the differential thermal analysis (DTA) in the temperature range of 400 – 1450 K. There was no structural transition in the DTA curves for both the alloys. The absence of (111) peak in the XRD pattern of CoFeMnAl indicates the presence of B2 disorder as shown by Alijani et al. [45]. Further studies are needed to explore the crystal structure of CoFeMnGa as no conclusions were made by them [45] on the basis of room temperature XRD data. The results obtained from various measurements are summarized in table 2.

**B. CoFeCrZ (Z = Al and Ga) alloys**

Bainsla *et* al. [9] found that CoFeCrAl alloy exists in B2-type cubic Heusler structure as revealed by the room temperature powder XRD analysis shown in Fig. 3(*c*). Absence of (111) in the XRD pattern indicates that the alloy exists in B2-type structure. Luo et al. [47] and Nehra et al. [48] also observed the B2-type structure for CoFeCrAl on the basis of XRD and Mössbauer analysis. The degree of long range B2 order ($S_{B2}$) was roughly estimated by Bainsla *et* al. [9] and it was found to be 0.89 from the intensity ratio of (200) peak and (400) fundamental peak [49]. Such a large value indicates the presence of a highly ordered B2 structure. The XRD results of Bainsla et al. are found to be in good agreement with the reports by Luo et al. [47] and Nehra et al. [48].

CoFeCrGa alloy is found to exist in the Y-type structure as revealed by the room temperature powder XRD {as given in Fig. 3(*d*)} and Mössbauer analysis [10]. The lattice parameter of this alloy is found to be 5.79 Å. The superlattice reflections [(111) and (200)] were found to absent in the XRD pattern due to the nearly equal scattering amplitudes of the Co, Fe, Cr and Ga. It is not necessary that the absence of SR in such cases is due to the presence of structural disorder, as shown above in the case of $Co_2FeZ$ (Z=Al, Si, Ga, Ge) alloys [42]. To further investigate the crystal structure of CoFeCrGa, Bainsla *et* al. [10] performed room temperature $^{57}Fe$ Mössbauer spectroscopic measurements. Mössbauer spectrum was best fitted with two sextets having hyperfine field values of 254 and 142 kOe and relative intensities of 58, 34 respectively, along with a doublet of 8% intensity (paramagnetic). In a well ordered Y-type structure there must be a single $H_{hf}$ due to the presence of only one crystallographic site for Fe. Thus the presence of second sextet indicates the structural disorder for the alloy. The authors thus concluded that CoFeCrGa alloy exists in Y-type crystal structure along with some $DO_3$ disorder. They also checked the phase stability of the crystal structure by performing DTA in the temperature range of 500 - 1600 K, a peak in the DTA curves was obtained around 1000 K, which was attributed to the ordered Y-type to the disordered B2 phase transition.

**C. CoRuFeZ (Z = Si and Ge) alloys**

The crystal structure of CoRuFeZ (X = Si and Ge) was investigated by Bainsla *et* al. [11] and Deka *et* al. [50] using XRD and Mössbauer analysis. (111) and (200) peaks were found to be present in the XRD for both the alloys [11], as shown in Fig. 3(*e*)-3(*f*). However, in the case of CoRuFeSi, the intensity of (200) peak was high, while (111) reflection was weak, giving an indication of B2 disorder in CoRuFeSi. However, the superlattice reflections were weak in the case of CoRuFeGe, which was attributed to the nearly same scattering amplitudes of all the sites. The lattice parameter values of 5.77 and 5.87 Å were obtained for CoRuFeSi and CoRuFeGe respectively.

Bainsla *et* al. [11] further investigated the crystal structure of these alloys by performing the room temperature $^{57}$Fe Mössbauer spectroscopic analysis. The best fit to the Mössbauer data was obtained with two sextets (S1 and S2) for both the alloys. From this, it was found that both the alloys are fully ferromagnetic at room temperature as no doublet was seen in the Mössbauer spectra. The cubic symmetry in the alloys was confirmed by almost vanishing values of quadruple shift. In the case of CoRuFeSi, $H_{hf}$ values of 298 and 314 kOe with relative intensities of 76 and 24% were obtained for S1 and S2 respectively, while in the case of CoRuFeGe, $H_{hf}$ values of 307 and 328 kOe with relative intensities of 88 and 12% were obtained for S1 and S2 respectively. The presence of two sextets for both the alloys indicates the presence of chemical disorder in the unit cell. The obtained $H_{hf}$ values from the fit suggest that Fe also occupies Z site, which results in the B2 type crystal structure. The intensities of S1 and S2 were attributed to the Y and B2 phases respectively for CoRuFeSi and CoRuFeGe. Both the alloys were found to be reasonably ordered at room temperature, as revealed by the relative intensities of S1 and S2. Therefore, it was concluded that the substitution of Ru in place of Mn in CoFeMnZ (Z = Si and Ge) [which results in CoRuFeZ] improves the chemical ordering in these alloys [11]. The results of the structural analysis for these alloys are given in the table 2.

**D. CoFeTiAl and CoMnVAl alloys**

Structural details of these alloys were obtained by Basit *et* al. [51] using the room temperature XRD analysis. According to them, the typical SR (111) and (200) were clearly seen in the XRD pattern of these alloys. The refinement of the XRD data revealed that these alloys exist in Y-type structure with lattice parameter values of 5.851 and 5.805 Å for CoFeTiAl and CoMnVAl respectively, as given in table 2. The difference between the measured data and the theoretical data for (200) peak indicated the presence of some amount of chemical disorder. They tried various combinations (disorder, interchange of different atoms) to improve the fitting between experimental and theoretical data in their refinement. However, the exact amount of structural disorder in these alloys has not been reported from the XRD analysis.

**E. CoFeCrGe and CoMnCrAl alloys**

Enamullah et al. [52] obtained the structural information of these alloys by performing the Rietveld refinement of the room temperature XRD data. As per their report, these alloys are found to exist in cubic Heusler structure with lattice parameters of 5.77 and 5.76 Å for CoFeCrGe and CoMnCrAl respectively (as given in table 2). The intensities of SR (111) and (200) were weak in the case of CoFeCrGe, which may be due to the nearly equal atomic scattering amplitudes of the atoms. According to these authors, the best fit between the experimental and calculated intensities was obtained for the configuration Ge and Cr at 4a(0, 0, 0) and 4b(1/2, 1/2, 1/2) octahedral sites respectively while Co and Fe at 4c(1/4, 1/4, 1/4) and 4d(3/4, 3/4, 3/4) tetrahedral sites respectively. *Ab-initio* calculations by the same researchers also predicted that the configuration which results in the best fit of the XRD data is indeed the energetically most favorable [52].

In the case of CoMnCrAl, the superlattice peak (200) was found to be more intense as compared to (111) peak as per the report of Enamullah et al. [52]. The observed XRD data clearly suggested a chemical disorder between the octahedral sites of the system. Similar to CoFeCrGe, there are possibilities of three different configurations for CoMnCrAl with the exception that in this case, exchange of atoms between octahedral sites. i.e. (0, 0, 0) and (1/2, 1/2, 1/2) fcc sublattices is also to be considered. Even though the observed XRD pattern fitted well with all the three configurations, the one in which the octahedral site (1/2, 1/2, 1/2) fcc containing the least electronegative element is found to be energetically the most favorable. Here Cr and Mn have the least electronegativity, and hence the two configurations which contain Cr or Mn at octahedral site are favorable. Electronegativity of the Al is also of the same order as of Mn and Cr, and thus Al tries to occupy (1/2, 1/2, 1/2) fcc sublattice in addition to the (0, 0, 0) sublattice. Due to this behavior of Al, the (111) peak is found to be absent from the XRD pattern.

Enamullah et al. [52] have also observed that if two or more atoms possess nearly same electronegativity values, some degree of disorder could be expected in the crystal structure. They gave the example of CoMnCrAl, CoFeCrAl [9] and $Co_2Cr_{1-x}Fe_xAl$ [53] alloys, where disorder is found between Cr and Al sites. On the basis of the data available in the literature, they proposed an empirical relation between the relative electronegativity values and the occurrence of disorder.

**F. NiCoMnZ (Z = Al, Ge and Sn) alloys**

Structural analysis of these alloys was done using powder XRD as well as neutron diffraction measurements by Halder et al. [54]. The Rietveld refinement of the room temperature powder XRD data revealed that all the alloys formed in single phase with cubic Heusler structure. The SR (111) and (200) were found to be absent for NiCoMnAl and NiCoMnGe, while they were present in NiCoMnSn. The appearance of the (111) and (200) peaks in the XRD pattern of NiCoMnSn indicates the absence of structural disorder. Halder et al. also mentioned that mere absence of the superlattice reflections in the XRD of NiCoMnAl and NiCoMnGe should not be taken as the sign of the disordered crystal structure. Thus, to further investigate the crystal structure of NiCoMnAl and NiCoMnGe, they performed neutron diffraction measurements at different temperatures. They analyzed the neutron diffraction pattern using the Rietveld refinement (using the FULLPROF program). The superlattice reflections (111) and (200) were observed in the neutron diffraction pattern of NiCoMnGe, in contrast to their XRD study. In the case of NiCoMnAl, they observed (200) peak in the neutron diffraction spectrum, while (111) peak was absent. The presence of the (200) peak in the neutron diffraction data, in contrast to the absence of both (111) and (200) reflections in the XRD data, shows the importance of the neutron diffraction as a probe in such cases [where the transition elements and the main group element belong to the same period]. Thus, Halder et al. [54] concluded that B2 type structural disorder is present in the case of NiCoMnAl.

In order to get further insight into the order/disorder for NiCoMnAl and NiCoMnGe alloys, Halder et al. [54] performed model calculations with varying amounts of disorder. They quantified the disorder with respect to the Mn-Z and Ni-Co sites. Due to the differences in the scattering lengths of neutrons for Ni and Co, they could identify the disorder at Ni and Co sites. Thus, on the basis of the observed neutron diffraction data, and model calculations, they concluded that there was a complete disorder at Mn-Al sites as well as at Ni-Co sites for NiCoMnAl. On the other hand, in the case of NiCoMnGe, they concluded that there was no disorder at Mn-Ge sites, but they showed a complete disorder at Ni-Co sites. From the temperature variation of neutron diffraction data, they found that the lattice parameters for both the alloys increased with increase in temperature. The results of structural analysis for NiCoMnZ (Z = Al, Ge and Sn) are also given in table 2.

### G. NiFeMnGa, NiCoMnGa and CuCoMnGa alloys

Alijani et al. [55] obtained the structural details of these alloys by performing the Rietveld refinement of the room temperature XRD patterns. All the alloys were found to exist in the cubic Y-type Heusler structure. The SR (111) and (200) were not observed in the XRD for these alloys, which is due to the nearly equal scattering amplitudes of all the elements. Thus, a conclusive structural analysis was not possible from the measurements, and a deeper structural analysis is required for these alloys. According to them, the determination of exact crystal structure in such cases is possible from anomalous XRD and EXAFS measurements. The lattice parameters obtained from the refinement of the XRD data are given in the table 2 for all the alloys. They also calculated the bulk moduli of all the alloys, and found that the Ni containing alloys are harder as compared to Cu containing alloy.

### H. MnNiCuSb alloy

The structural properties of MnNiCuSb were investigated by Haque *et al* [56] using the powder x-ray diffraction measurements. This alloy was found to exist in single phase with cubic Heusler crystal structure (Y-type) and $F\bar{4}3m$ space group (#216). However, small Cu impurity peak (< 5%) was observed in the XRD pattern. Lattice parameter value of 5.923 Å was obtained by performing the Rietveld refinement of XRD data, as given in the table 2. The superlattice reflections (111) and (200) were present in the XRD with moderate intensity, which hint at the ordered crystal structure. However, as mentioned earlier, it is very difficult to probe the structural disorder in the systems where all the transition elements belong to the same period in the periodic table. Elemental composition of the alloy was also studied by Haque *et al* [56], using the energy dispersive X-ray spectroscopy (EDS) studies. The final stoichiometry of the alloy was calculated by averaging the data obtained over many grains and found to be $Mn_{0.97}NiCu_{0.95}Sb$, which is very close to the 1:1:1:1 composition. The homogeneity of the alloy was also confirmed by the EDS analysis.

## III. MAGNETIC PROPERTIES

Unlike the usual ferromagnets, HMFs exhibits integer magnetic moments, usually obeying the modified Slater-Pauling rule [57-59]. The magnetic moment in Heuler alloys, especially in $Co_2$ based alloys follows the Slater-Pauling rule [57-59]. In the case of localized moment systems, an average magnetic moment ($m_{av}$) per atom can be defined as [55],

$$m_{av} = (n_{av} - 6 - 2n_{sp})\mu_B \quad (1)$$

Here, $n_{av}$ is the average number of valence electrons per atom in the alloy and $n_{sp}$ represents the average number of unbalanced sp electrons. The number 6 arises because Fermi level lies in the gap between of the occupied and the unoccupied d electron states for spin down states. In half-metallic ferromagnetic systems, where a band gap is present in one of the spin sub bands and all *sp* electrons are occupied, $n_{sp}$ vanishes. There are four atoms in the primitive cell of 2:1:1 and 1:1:1:1 Heusler alloys. Thus, the total magnetic moment (*m*) in the case of 2:1:1 and 1:1:1:1 half-metallic Heusler alloys is [57-59],

$$m = (N_v - 24)\mu_B \quad (2)$$

where, $N_v$ is the total number of valence electrons in the primitive cell (see details in reference 58). The total number of valence electrons in a 2:1:1 and 1:1:1:1 Heusler alloy system, can be obtained using the relation,

$N_v$ = [(outer s and d electrons for transition metals) + [(outer s and p electrons for main group elements)] (3)

In the case of non half-metallic state, the magnetic moment value will deviate from the equation (2). According to the Slater-Pauling rule, ferromagnetic ordering is found to be absent at $N_v$=24, due to a quasi-closed-shell character. The phenomenon of half-metallic, fully compensated ferrimagnetism was predicted theoretically in certain cases of Heusler alloys with 24 valence electrons [60]. In the case of $N_v \geq 30$, the assumption of localized magnetic behavior is not valid. Heusler alloys with magnetic moment values greater than 6 $\mu_B$ are rarely known. Thus in the case of high valence electron concentration ($N_v \geq 30$), an itinerant Slater-Pauling behavior is approximated [61,62]. The following relation is valid for such cases:

$$m = (34 - N_v)\mu \quad (4)$$

## A. CoFeMnZ (Z = Al, Si, Ga and Ge) alloys

Magnetic properties of these alloys were first investigated by Alijani et al. [45] using a SQUID magnetometer. The corresponding saturation magnetization ($M_S$) and the Curie temperature values are given in table 3. They observed that the $M_S$ value for all the alloys werehigher in comparison to the values obtained using Slater-Pauling rule [57-59]. They attributed the difference between the experimental and theoretical results to the presence of disorder and/or magnetic impurities. As discussed above (structural properties section), structural disorder was observed in the case of CoFeMnAl and CoFeMnSi, while Alijani et al. [45] could not obtain any structural disorder for CoFeMnGa and CoFeMnGe. However, they were not able to identify the impurities from the XRD analysis for both CoFeMnGa and CoFeMnGe. Bainsla et al. [7,8] further investigated the magnetic properties of CoFeMnZ (Z = Si and Ge) alloys, and they obtained the values of 3.7 and 4.2 $\mu_B$/f.u. at 3 K for CoFeMnSi and CoFeMnGe respectively as shown in the Fig. 4(*a*). The $M_S$ values of 4.1 and 4.2 $\mu_B$/f.u. at 5 K were obtained for CoFeMnSi and CoFeMnGe respectively by Alijani et al. [45]. The difference between the $M_S$ values for CoFeMnSi could be attributed to the large structural disorder observed by Bainsla et al. [8,46].

The Curie temperatures of all the alloys were first determined using the temperature dependent magnetization measurements by Alijani et al. [45] and $T_C$ was found to be much above room temperature for all of them, as given in table 3. High values of $T_C$ facilitates in technological applications of these materials. Bainsla et al. [7,8] used differential thermal analysis to estimate the Curie temperature of CoFeMnZ (Z = Si and Ge) alloys and obtained values close to 620 and 750 K, in comparison to the 623 and 711 K (obtained by Alijani et al., reference 45) for CoFeMnSi and CoFeMnGe respectively.

In order to understand the magnetic properties further, the magnetic behavior of these alloys were studied using *ab-inito* calculations by Alijani et al. [45]. The calculated values of $M_S$ were found to be integer, typical for half-metallic ferromagnets. As per the Slate-Pauling rule, $M_S$ adopts values of 3 $\mu_B$/f.u. for Z = Al and Ga, while 4 $\mu_B$/f.u. for Z = Si and Ge. The site specific magnetic moment calculations by Alijani et al. revealed that the magnetic properties are mainly determined by the Mn atoms, which contribute the highest magnetic moment (~2.5 $\mu_B$). They also found that the moment of Co sites is stable against the variation of the Z element or the number of valence electrons, and exhibits a value of ~0.8 $\mu_B$. The most interesting observation of these authors is that the moment at Fe sites strongly varied with the number of valance electrons and it even changed sign as the number of valence electrons changed from 27 to 28. In CoFeMnAl and CoFeMnGa alloys, the Fe atoms were found to be in antiparallel alignment with Co and Mn atoms. *i.e.* in ferrimagnetic order. In a Y-type structure, the nearest neighbors of Mn are Co and Fe. Thus it can be supposed that magnetic moment at Fe sites is induced by the neighboring Mn spins. Alijani et al. [45] observed that the stability of the Co and Mn moments together with the Slater-Pauling rule dictates whether the moment at Fe sites are aligned parallel or antiparallel to Mn.

Further investigation on the local magnetic moment and other magnetic properties were made by Klaer et al. [63] using the circular dichroism in x-ray-absorption spectroscopy (XAS/XMCD). They performed XAS/XMCD measurements at the Mn, Fe and Co $L_{3,2}$ edges. The shape of the spectra was found to be pairwise similar for CoFeMnAl(Ga) and for CoFeMnSi(Ge), which was expected due to the same crystal structure and isoelectronic nature of the pairs. According to them, the largest change in XMCD spectra was observed for Mn. The magnetic moment at Mn site changes from 0.96 $\mu_B$ to 2.22 $\mu_B$ for Z = Al to Z = Si respectively. The maxima at $L_{3,2}$ edge appear much sharper for Co and Fe for Z= Si and Ge, in comparison to Z = Al and Ga. The variation of Z in CoFeMnZ alloys shows no significant effect on the Co and Fe moments, but in contrast, the increase in the number of valence electrons causes a considerable increase in the Mn moment. The total moment values derived from SQUID measurements (as performed by Alijani et al., reference 45) were found to be in good agreement with those obtained using XMCD analysis by Klaer et al. [63].

As discussed above, depending on the occupancy of different lattice sites, three different structural configurations are possible for Y-type structure. Klaer et al. [63] tried to find out the correct crystal structure for these alloys by comparing the theoretically obtained individual magnetic moment values with those obtained using the XMCD analysis. They found that theoretically obtained crystal structure is in agreement with XMCD analysis for Z = Si and Ge, while, element specific magnetic moment observations using XMCD ruled out the theoretical predictions for Z = Al and Ga.

**B. CoFeCrZ (Z = Al and Ga) alloys**

Magnetic properties of CoFeCrAl alloy were first measured by Luo et al. [47]. Curie temperature of 460 K for the alloy was obtained using the thermomagnetic curves recorded by them. Nehra et al. [48] also studied the magnetic properties of the as-melted as well as quenched CoFeCrAl alloy and found a bifurcation between the zero field cooled and field cooled thermomagnetic data. They also observed that the difference between these two data decreases with increase in the cooling field and entirely disappears in a cooling field of ~1.5 kOe. These authors attributed the magnetic behavior to the antiferromagnetic interactions between Cr-Co/Fe and Cr at ordinary sites with Cr at the anti (Al) sites. The $T_C$ values of 410 and 515 K were obtained for as-melted and quenched CoFeCrAl samples respectively [48]. Bainsla et al. [9] further investigated the magnetic properties of CoFeCrAl alloy and found that $T_C$ value was more than 400 K.

The saturation magnetization value of 2.07 $\mu_B$/f.u. was obtained for CoFeCrAl at 5 K by Luo et al. [47], which was found to be in agreement with their theoretically predicted value of 2.0 $\mu_B$/f.u. Bainsla et al. [9] also investigated the magnetic properties of CoFeCrAl alloy by performing the pressure dependent isothermal magnetization measurements at 8 K. The $M_S$ value of ~2.0 $\mu_B$/f.u. was obtained under the ambient pressure as shown in the Fig. 4(*c*) and given in the table 3, and it is found to be invariant under pressure up to 12 kbar. The invariance of $M_S$ was observed earlier in the case of Co$_2$VGa [64] and it was attributed as a characteristic of half-metallic ferromagnets. The theoretically predicted $M_S$ value (1.97 $\mu_B$/f.u.) by Bainsla et al. [9] is found to be in good agreement with the value (2.0 $\mu_B$/f.u.) obtained using Slater-Pauling rule [57-59]. The effect of lattice compression in certain Heusler alloys was studied by Picozzi et al. [38] using ab-initio calculations, and they found that the half-metallic nature can be preserved under lattice parameter variation of 3-5%.

Bainsla et al. [10], studied the magnetic properties of CoFeCrGa alloy by recording the thermomagnetic curves in the temperature range of 5-400 K with applied magnetic field of 500 Oe. Curie temperature value is expected to be more than 400 K as no magnetic transitions were observed in the thermomagnetic curves up to that temperature. The isothermal magnetization curves were recorded with different values of applied pressure (up to 9.8 kbar) at 8 K. The pressure independent $M_S$ value of ~2.1 $\mu_B$/f.u. was obtained and it was found to be in good agreement with the estimated value (2.0 $\mu_B$/f.u.) using Slater-Pauling rule. According to this rule, integer values of the total moment are possible in the case of the half-metallic ferromagnetic systems. The isothermal magnetization curve at 8 K and under ambient pressure for CoFeCrGa is given in Fig. 4(*c*). As discussed above, the pressure independent $M_S$ value is a sign of half-metallic ferromagnetic nature in CoFeCrGa.

Bainsla et al. [10], also performed *ab-initio* calculations to get better insight into the magnetic behavior of CoFeCrGa alloy. As per these calculations, the total magnetic moment was found to be constant as a function of lattice parameter, which supports their experimental claims. The Fe moment was found to be aligned antiferromagnetically to Co and Cr for all lattice parameter values. The Co moment remains constant under the application of hydrostatic pressure, while it causes a decrease in the moments of antiferromagnetically coupled Fe and Cr, and collectively results in a constant total magnetic moment.

### C. CoRuFeZ (Z = Si and Ge) alloys

Magnetic properties of CoRuFeSi alloy were first studied by Deka et al. [50], using a vibrating sample magnetometer. The $M_S$ value of ~5.0 µ$_B$/f.u. was obtained at 300 K by them for CoRuFeSi. Bainsla et al. [11], investigated the magnetic properties of the CoRuFeZ (Z = Si and Ge) alloys by studying the isothermal magnetization curves at different temperatures and thermomagnetic curves up to 1100 K. The saturation magnetization values of 4.8 and 5.0 µ$_B$/f.u. at 5 K were obtained for CoRuFeSi and CoRuFeGe respectively, as shown in the Fig. 4(*b*) and table 3. The $M_S$ values are found to be slightly different at room temperature (300 K), 4.6 and 4.7 µ$_B$/f.u. for CoRuFeSi and CoRuFeGe respectively. The obtained $M_S$ values at 5 K are found to be in good agreement with those estimated using the Slater-Pauling rule [57-59]. Thus, the moment value of 5.0 µ$_B$/f.u. (5 K) obtained for CoRuFeGe gives a strong indication of the half-metallic nature.

The Curie temperature of the alloys were determined by Bainsla et al. [11] by recording the thermomagnetic curves under a field of 50 Oe in the range of 300-1100 K as shown in Fig. 5(*a*) – 5(*b*). $T_C$ values of 867 and 833 K were obtained for CoRuFeSi and CoRuFeGe respectively, in which the former is found to be the highest among all the known equiatomic quaternary Heusler alloys [7-11,45,46,52,63]. The relatively high value of $T_C$ in these alloys gives a possibility of retaining high spin polarization at elevated temperatures.

### D. CoFeTiAl and CoMnVAl alloys

The magnetic properties for these alloys were studied by Basit *et* al. [51] by measuring the magnetization isotherms at different temperatures for CoFe$_{1+x}$Ti$_{1-x}$Al and CoMn$_{1+x}$V$_{1-x}$Al alloys. These authors observed that the parent alloys namely CoFeTiAl and CoMnVAl do not show any moment at 5 K, which is in agreement with the Slater-Pauling rule. Due to the better structural stability in CoMn$_{1+x}$V$_{1-x}$Al alloys as compared to CoFe$_{1+x}$Ti$_{1-x}$Al alloys, they performed *ab-initio* calculations in CoMn$_{1+x}$V$_{1-x}$Al alloys to understand the electronic structure. The absence of magnetic moment in the parent CoMnVAl was already expected from the Slater-Pauling rule. The experiments performed by them did not give any information on the magnetic order of CoMnVAl. However the possibility of a fully compensated half-metallic ferrimagnetism could not be ruled out [60,65], with a ferromagnetic-paramagnetic transition taking place at temperatures below 5 K. Ab-initio calculations have shown a non spin-polarized ground state for CoMnVAl and it was obtained independently of the method using starting parameters involving ferromagnetic or different types of ferrimagnetic order. Therefore, fully compensated half-metallic ferrrimagnetic ground state can be excluded as the ground state.

**E. CoFeCrGe and CoMnCrAl alloys**

The magnetic properties of these alloys were studied by Enamullah et al. [52], using a Physical Property Measurement System. Enamullah et al. [52], recorded thermomagnetic curves under a field of 500 Oe for both the alloys as shown in the Fig. 5(*c*) – 5(*d*). The values of 358 and 866 K were obtained for CoMnCrAl and CoFeCrGe respectively. The $T_C$ value estimated for the CoFeCrGe alloy is found to be nearly equal to the highest $T_C$ (867 K) value which was obtained for CoRuFeSi.

Enamullah et al. [52], also recorded isothermal magnetization curves at different temperatures. The soft ferromagnetic nature was observed for these alloys, as there was no hysteresis found in the *M-H* curves. Both the alloys shows saturation in magnetization at 5 K {as shown in Fig. 4(*d*)} and 300 K. The saturation magnetization values of 3.0 and 0.9 µ$_B$/f.u. were obtained at 5 K for CoFeCrGe and CoMnCrAl respectively, while the $M_S$ values of 3.0 and 1.0 µ$_B$/f.u. were estimated using the Slater-Pauling rule [57-59]. The presence of structural disorder in CoMnCrAl might be the possible reason for the deviation in $M_S$ value. *Ab-inito* calculations were also performed by them on these alloys [52]. The calculated magnetic moment values of ~1.0 and ~3.0 µ$_B$/f.u. were found for CoMnCrAl and CoFeCrGe respectively, which follow the Slater-Pauling rule.

**F. NiCoMnZ (Z = Al, Ge and Sn) alloys**

Halder *et* al. [54], studied the magnetic properties of these alloys using magnetization and neutron diffraction measurements. Thermomagnetic curves were obtained under an applied field of 500 Oe for all the alloys and the $T_C$ values were found to be higher than 583 K. Isothermal magnetization curves were obtained at different temperatures for all the alloys over a field range of 50 kOe by them. The *M vs. H* curves at 5 K showed a soft ferromagnetic nature as revealed by negligible hysteresis. The obtained $M_S$ values at 5 K and corresponding magnetic moment values from *ab-initio* calculations for all the alloys are given in table 3. The $M_S$ values estimated using the Slater-Pauling rule are found to be 5.0, 6.0 and 6.0 µ$_B$/f.u. for NiCoMnAl, NiCoMnGe and NiCoMnSn alloys respectively, in comparison to the experimentally obtained values of 4.7, 4.1 and 4.8 µ$_B$/f.u. The *ab-initio* calculation predicted that the half-metallic nature was absent for both NiCoMnGe and NiCoMnSn, and the magnetic moment was less than 6.0 µ$_B$/f.u. According to this study, one of the possible reasons for the absence of half-metallic nature is that the Ni remains almost paramagnetic in Ni based quaternary Heusler alloys and cannot carry a large magnetic moment. The discrepancy between the experimentally obtained $M_S$ values and the calculated magnetic moment values was attributed to the presence of structural disorder as discussed in section 2.6.

Halder *et* al. [54] further investigated the magnetic properties of these alloys by performing the neutron diffraction measurements for NiCoMnAl and NiCoMnGe alloys. The analysis of the neutron diffraction data reveals the ferromagnetic ordering of Co and Mn moments for both the alloys. However, no ordered moment of the Ni atom could be refined satisfactory in both the alloys. The magnetic moment values of 0.97(6) and 2.95(4) µ$_B$/f.u. were obtained for Co and Mn respectively at 6 K in the case of NiCoMnGe. In the case of NiCoMnAl, moment values of 1.08(9) and 3.13(7) µ$_B$/f.u. were obtained for Co and Mn respectively at 6 K.

**G. NiFeMnGa, NiCoMnGa and CuCoMnGa alloys**

The magnetic properties of these alloys were studied by Alijani *et* al. [55] using the SQUID magnetometer. Isothermal magnetization curves show negligible hysteresis, revealing the soft ferromagnetic nature. The $M_S$ and $T_C$ values of these alloys are summarized in the table 3. The $M_S$ values of 3.4, 4.5 and 2.3 µ$_B$/f.u. were obtained at 5 K for NiFeMnGa, NiCoMnGa and CuCoMnGa alloys respectively. According the Slater-Pauling rule [57-59] and *ab-initio* calculations by Alijani *et* al., the moment values of 4.0 and 5.0 µ$_B$/f.u. were obtained for NiFeMnGa and NiCoMnGa respectively. As can be noted, there is a large discrepancy between experimentally obtained $M_S$ values and calculated values, which could be attributed to the structural disorder. But these authors observed that the difference was too high to be dedicated only to the disorder effect. Impurities below the detection limit of XRD (approximately 5%) may also contribute to this difference. They also mentioned the possibility of a canted spin structure, which was found energetically unstable for these alloys as per their calculations.

The total moment of CuCoMnGa was estimated to be 6.0 $\mu_B$/f.u. when assuming half-metallicity and localized Slater-Pauling behavior [57-59], which was found to be in disagreement with experiment [55]. The magnetic moment of 4.0 $\mu_B$/f.u. was estimated when considering an itinerant Slater-Pauling behavior [55,61,62]. The experimentally obtained low value of moment in the case of CuCoMnGa was explained by a vanishing contribution of Cu to the magnetic properties. Although, the results of *ab-initio* calculations predict absence of half-metallic nature, the experimental $M_S$ value is so small to compare in the case of CuCoMnGa.

According to Alijani *et* al. [55], $M_S$ values for NiFeMnGa and NiCoMnGa show fair agreement with the Slater-Pauling rule [57-59] and a possible half-metallic ferromagnetic behavior, making them suitable for spintronic applications. $T_C$ values of 326, 646 and 631 K were obtained for NiFeMnGa, NiCoMnGa and CuCoMnGa respectively. The relatively high $T_C$ in half-metallic NiCoMnGa makes it even more attractive for the applications. Thermomagnetic curves recorded by Alijani *et* al. [55] show the conventional ferromagnetic behavior for NiCoMnGa and CuCoMnGa, while NiFeMnGa shows a sharp drop in magnetization around $T_C$.

**H. MnNiCuSb alloy**

Haque et al. [56], investigated the magnetic properties of MnNiCuSb using the SQUID magnetometer. Thermomagnetic curve was recorded in the temperature range of 300 – 1000 K, and $T_C$ = 690 K was obtained. Isothermal magnetization curve at 5 K shows negligible hysteresis and thus a soft ferromagnetic nature. The $M_S$ value of 3.85 $\mu_B$/f.u. was obtained at 5 K using the *M vs. H* curve.

**IV. MAGNETO-TRANSPORT PROPERTIES**

Electrical resistivity ($\rho$) is an intrinsic property of a solid, which quantifies how strongly it opposes the flow of electric current. Thus, $\rho$ describes the response of a system due to an external electric field. These measurements are very effective and useful in probing the magnetic state of the materials because of the strong coupling between the electron mobility and the magnetic state. The origin of electrical resistivity in metals and alloys is the scattering of charge carriers *viz.* the conduction electrons by lattice defects, thermally excited lattice vibrations (phonons), spin waves (magnons), spin fluctuations etc. Another contribution arises from the electron-electron interaction, mainly at low temperatures. The conduction band in these materials is mainly constituted by the s-electrons. Due to their large effective mass, the contribution to the conductivity from the d-electrons is quite negligible. However, d-electrons play an important role in determining the resistivity and its variation due to the large density of states of the d-band. This is due to the fact that scattering probability depends on the DOS into which the electrons are scattered. Therefore, s-d scattering plays an important role in the electrical resistivity behavior in metallic systems. In the case of magnetically ordered systems, the change in the long-range order is reflected in $\rho$ behavior and often manifests as a change in the slope of a $\rho$ *vs.* $T$ curve at the transition temperatures.

The total resistivity of magnetic alloys is generally found to obey the Matthiessen's rule [66]. According to this rule, the contributions from different factors to the total resistivity are independent of one another and thus additive. The total resistivity can be written as:

$$\rho(T) = \rho_0 + \rho_{ph}(T) + \rho_{mag}(T) \tag{5}$$

Here $\rho_0$ is the residual resistivity, which is temperature independent and originates from the scattering of conduction electrons by the lattice defects, impurities etc. $\rho_{ph}$ and $\rho_{mag}$ arise due to the scattering by phonons and magnons respectively and are temperature dependent. The phononic part of the scattering process ($\rho_{ph}$) can be described by the Bloch-Grüneisen relation [67],

$$\rho_{ph} = 4R\left(\frac{T}{\theta_D}\right)^5 \int_0^{\theta_D/T} \frac{x^5}{(1-e^{-x})(e^{-x}-1)} dx \tag{6}$$

Where R includes the electron-phonon coupling constant and atomic masses of different type of atoms, and $\theta_D$ is the Debye temperature. In the high temperature limit, *i.e.* T > $\theta_D$, $\rho_{ph}$ varies linearly with temperature. For the low temperature limit *i.e.* T < $\theta_D$, $\rho_{ph}$ shows a $T^5$ dependence.

At low temperatures, the electrical resistivity of a material originates from the (i) electron-electron scattering [68], (ii) electron-magnon scattering [69,70] and (iii) electron-phonon scattering [68]. The electron-phonon contribution is relatively smaller at low temperatures and thus $\rho$ is mainly contributed by the electron-electron and electron-magnon scatterings. In the case of ideal HMFs, electron-magnon scattering is found to be absent due to unavailability of the minority carriers at the $E_F$. As the temperature increases, the half metallic trend decreases and the electron-magnon contribution appears. However, a double magnon process is possible in HMFs. Therefore, the major contributions to the resistivity in a HMF arise due to (i) electron-electron ($T^2$ dependence) [68], (ii) electron-phonon (T dependence) [68], and (iii) double magnon scattering ($T^{9/2}$ at low temperatures and $T^{7/2}$ at high temperatures) [71,72].

The change in the resistivity on the application of an external magnetic field is known as magnetoresistance (MR), which can be used as a probe to study magnetic systems. MR can be estimated from the field dependence of the resistivity data and is given by the relation,

$$MR\% = \left(\frac{\rho(T,H) - \rho(T,0)}{\rho(T,0)}\right) \times 100 \qquad (7)$$

Where $\rho(T,H)$ and $\rho(T,0)$ are the resistivity values of the material at temperature $T$ under the application of an applied magnetic $H$ and in the absence of the field respectively. In any material, the application of a magnetic field deviate conduction electrons from their path and thus leads to an increase in the resistivity value, thereby a positive MR. The positive increase in the MR is termed as Lorentz contribution [73]. In spite of the presence of the Lorentz contribution, intermetallic compounds in general show a decrease in the resistivity, thereby resulting in negative MR, which may originate from the suppression of the spin fluctuations by the field or by the decrease in the spin disorder scattering. The negative MR arising from the suppression of the spin fluctuations follows quadratic field dependence in the paramagnetic region [74]. Positive MR is obtained in many antiferromagnetic materials, which originates from the enhancement of spin fluctuations by the application of a magnetic field. The change in the magnetic structure of materials, especially in antiferromagnets, often leads to the creation of a superzones in their band structure and thus gives rise to positive MR [75]. The HMF materials generally show low MR values due to the absence of spin fluctuations [9,76-78].

In the following subsections, a brief discussion on the various magneto-transport measurements reported for some EQHAs is given.

## A. CoFeMnZ (Z = Si and Ge) alloys

Bainsla *et* al. [8] studied the temperature dependencies of the electrical conductivity [$\sigma_{xx}(T)$] and the carrier concentration [$n(T)$] in case of CoFeMnSi, the curves are shown in Fig. 6(*a*)-6(*b*). $\sigma_{xx}(T)$ was measured under different applied magnetic field (0 and 50 kOe) in the temperature range of 5-300 K. Electrical conductivity increases with increase in temperature, indicating a non-metallic conduction. $\sigma_{xx}(T)$ = 2980 and 3000 S/cm were obtained at 300 K under the fields of 0 and 50 kOe, respectively. The obtained electrical conductivity value at 300 K was found to be slightly higher than that observed for $Mn_2CoAl$ (i.e. 2440 S/cm) reported by Quardi *et* al. [26]. Linear behavior in $\sigma_{xx}(T)$ was observed in the high temperature region, while a non-linear behavior was observed at low temperatures for CoFeMnSi. The non-linear behavior was attributed to the disorder enhanced coherent scattering of conduction electrons [79]. The conductivity behavior for CoFeMnSi is reported to be unusual and different from that of the normal metals or semiconductors. The carrier concentration (*n*) value of the CoFeMnSi was calculated using the Hall coefficient ($R_H$) measurements, and a nearly temperature independent carrier concentration was obtained in the range of 5 – 300 K which is typical of spin gapless semiconductors [26,80]. The carrier concentration value of 4 x $10^{19}$ $cm^{-3}$ was observed at 300 K for CoFeMnSi, which is comparable to that observed for HgCdTe ($10^{15}$ – $10^{17}$ $cm^{-3}$), $Mn_2CoAl$ ($10^{17}$ $cm^{-3}$) and $Fe_2VAl$ ($10^{21}$ $cm^{-3}$) [81]. It may be noted that the physical origin behind such a temperature dependence in carrier concentration of gapless semiconductors are well established [80,82]. The carrier concentration in the case of gapless semiconductors could be written as [80]

$$n_i = p_i = 2m_e m_h^{1/2} \left[ \frac{k_B T}{2\pi \hbar^2} \right]^{3/2} \quad (8)$$

where $m_e$ and $m_h$ are the effective masses of the electrons and holes. A good fit to this expression revealed the spin gapless semiconducting behavior in CoFeMnSi [8]. Thus, the behavior of $\sigma_{xx}(T)$ and $n(T)$ strongly supports the SGS behavior in CoFeMnSi. The observations of low carrier concentration and high resistivity show the exceptional stability of electronic structure and its insensitivity to the structural disorder in CoFeMnSi. According to Bainsla *et* al. [8], this is the first EQHA material to show SGS behavior.

Bainsla *et* al. [8] further investigated CoFeMnSi by calculating the anomalous Hall conductivity $\sigma_{xy} = \rho_{xy}/\rho_{xx}^2$ at 5 K, which was obtained from the magnetic field dependent transport measurements in order to study the low field behavior in more detail. Hall conductivity behavior was found to follow the same behavior as that of magnetization isotherm. The anomalous Hall conductivity ($\sigma_{xy0}$) value of 162 S/cm was obtained, which is defined as the difference in $\sigma_{xy}$ values at zero and the saturation fields. The obtained ($\sigma_{xy0}$) value for CoFeMnSi was found to be higher than that observed in $Mn_2CoAl$ (22 S/cm) [26], but less than that of the half metallic $Co_2FeSi$ (≈ 200 S/cm at 300 K) [69], and $Co_2MnAl$ (≈ 2000 S/cm) [83].

Electrical resistivity measurements were done under the magnetic field of 0 and 50 kOe in the range of 5- 300 K for CoFeMnGe by Bainsla *et* al. [7], as shown in Fig. 6(*c*). Resistivity value of 3.1 μΩm was found for CoFeMnGe, which is relatively high as compared to that of $Co_2Fe(Ga_{0.5}Ge_{0.5})$ (0.6 μΩm) alloy [36]. They attributed this to the presence of larger $DO_3$ disorder in CoFeMnGe. The residual resistivity ratio [$\rho_{300K}/\rho_{5K}$] was found to be low (1.02), which confirms the large disorder in the alloy. A minimum in the resistivity curve was found near 40 K, which was followed by an upturn at lower temperatures. Such a behavior has been reported in many Heusler alloys [84] and is typically attributed to the disorder enhanced coherent scattering of conduction electrons [79]. It was found that for some alloys, such a minimum has been explained on the basis of Kondo effect [85]. Resistivity behavior for CoFeMnGe was analyzed in the temperature range of 5 - 300 K, considering different scattering mechanisms [7]. In addition to the contributions mentioned above, a term representing the structural disorder with a temperature dependence of $T^{1/2}$ is also present in systems with non-negligible disorder [84]. Keeping all the above points in mind, Bainsla *et* al. [7] analyzed the resistivity data by performing the fitting in two different temperature regions, i.e., between 5 and 85 K and between 85 and 300 K. The resistivity trend was found to fit well with the relation, $\rho(T) = \rho_0 - AT^{1/2} + BT^2 + CT^{9/2}$ in the low temperature region, while the best fit is given by $\rho(T) = \rho_0 + DT + ET^{7/2}$ in the high temperature region. $T^{1/2}$ term was found to be dominating over the other two contributions in the low temperature region, which may be due to the presence of disorder. Thus, the minimum seen in the ρ (T) plot was attributed to the structural disorder. In the high temperature region, the linear dependence due to the electron-phonon contribution dominates, but along with a weak negative contribution from the $T^{7/2}$ term. Therefore, the resistivity data analysis for CoFeMnGe shows the possibility of half-metallic ferromagnetic nature even at elevated temperatures.

## B. CoFeCrZ (Z = Al and Ga) alloys

Bainsla *et* al. [9] obtained the $\rho(T)$ data for CoFeCrAl alloy using four probe technique, which shows a linear behavior up to 150 K as shown in Fig. 6(*c*) and is described very well by the relation $\rho(T) = \rho_0 + a*T$. Here, the first and second terms represent residual resistivity and the electron-phonon interaction term, respectively. Electrical resistivity behavior of this kind has been observed earlier for half-metallic ferromagnetic NiMnSb and PtMnSb at low temperatures (up to nearly 20 K) [86]. In the case of CoFeCrAl, $\rho(T)$ follows a linear behavior up to relatively higher temperatures (up to 150 K), indicating the possibility of half-metallic nature at temperatures as high as 150 K. $\rho(T)$ was fitted well with the relation $\rho(T) = \rho_0 + b*T^n$, with $n = 0.2$ in the range of 150 - 300 K (region 2). The $T^2$ term arising due to the electron-magnon interaction was found to be absent for $\rho(T)$ up to 300 K by Bainsla *et* al. [9], implying the absence of minority states at the Fermi level. The *ab-initio* calculations [9], which found a large half-metallic band gap of 0.31 eV, also support the statement derived from the temperature dependence of resistivity data. Bainsla et al. [9], also obtained a non-saturating magnetoresistance with field for CoFeCrAl, with a value of -1.6% at 5 K for 50 kOe.

The temperature dependencies of electrical conductivity ($\sigma_{xx}$) with zero applied field and carrier concentration [$n(T)$] for CoFeCrGa are shown in Fig. 6(*a*)-6(*b*). A non metallic conduction behavior was observed for $\sigma_{xx}$, which increases with increase in the temperature. The obtained value of $\sigma_{xx} = 3233$ S/cm at 300 K was found to be slightly higher than that observed for other SGS Heusler alloys, e.g. $Mn_2CoAl = 2440$ S/cm [26] and CoFeMnSi = 3000 S/cm [8]. The carrier concentration determined from the Hall coefficient measurements in the range of 5 - 280 K was found to be of the order of $10^{20}$ cm$^{-3}$ and nearly temperature independent up to 250 K, above which it increases abruptly. Bainsla *et* al. [10], found that the observations in CoFeCrGa are more or less identical to those of CoFeMnSi, except for the abrupt jump in the $n(T)$. They mentioned that the abrupt change in $n(T)$ above 250 K may arise due to the onset of thermal excitations that dominate over the half metallic (minority) band gap. Thus, a clear evidence of SGS behavior was seen in CoFeCrGa. They [10] also obtained the anomalous Hall conductivity ($\sigma_{xy}$) at 5 K using the magnetic-field-dependent transport measurements. $\sigma_{xy}$ (= $\rho_{xy}/\rho_{xx}^2$) value obtained for CoFeCrGa was found to be comparable to that obtained for other half-metallic Heusler alloys and it follows the same behavior as the magnetization isotherms. The anomalous Hall conductivity was calculated as the difference in $\sigma_{xy}$ values in zero and the saturation fields and it attains a value of 185 S/cm, which is quite identical to that obtained in CoFeMnSi [8].

## C. CoRuFeZ (Z = Si and Ge) alloys

Electrical resistivity measurements for CoRuFeZ (Z = Si and Ge) alloys were performed under the applied field of 0 and 50 kOe in the range of 5 – 300 K [11] and the curves under zero field are given in Fig. 6(*d*). The value of $\rho$ is found to be almost constant with the variation of field for both the alloys. Resistivity fittings were performed in two different temperature regimes, low temperature curves were fitted using $\rho(T) = \rho_0 + a \times T^n$, while $\rho(T) = \rho_0 + b \times T + c \times T^2 = \rho_0 + \rho_{ph.} + \rho_{mag.}$ were fitted to the high temperature region. At low temperatures (5 - 50 K), the resistivity was found to be independent of the temperature for CoRuFeSi, thus indicating the absence of spin flip scattering in that region. Such a behavior was observed earlier for $Co_2FeSi$, which was proved to be half-metallic ferromagnetic in nature up to a temperature of 103 K (band gap = 8.9 meV) [69]. Thus, at low temperatures, a perfect half-metallic ferromagnetic nature is expected for CoRuFeSi (up to 50 K). An indirect evidence of the half-metallic nature was also seen for CoRuFeGe at low temperatures [11]. In the high temperature region, the fitting of the resistivity curves indicates the absence of half-metallic nature [11].

## V. SPIN POLARIZATION

As mentioned earlier, in the case of HMF materials, one of the spin bands is metallic-like, while the other is semiconductor-like. This results in a ferromagnet with electrons of only one spin direction at the Fermi level and gives rise to highly spin polarized currents, useful for spintronic devices. In general, the spin polarization of a material is defined as

$$P = \frac{n_\uparrow(E_F) - n_\downarrow(E_F)}{n_\uparrow(E_F) + n_\downarrow(E_F)} \qquad (9)$$

where $n_{\uparrow(\downarrow)}(E_F)$ is the number of up(down) spin carriers at the Fermi level. Therefore, half metallic ferromagnets possess a conducting channel for one spin and band gap for the other spin. The intrinsic spin polarization of a material is defined by the equation 9, while the transport spin polarization ($P_C$) can be expressed as the imbalance in the majority and minority spin currents [87,88];

$$P_C = \frac{n_\uparrow(E_F)v_{F\uparrow} - n_\downarrow(E_F)v_{F\downarrow}}{n_\uparrow(E_F)v_{F\uparrow} + n_\downarrow(E_F)v_{F\downarrow}} \qquad (10)$$

where $n_{\uparrow(\downarrow)}(E_F)$ and $v_{F\uparrow(\downarrow)}$ are the DOS at the Fermi level and Fermi velocity for up(down) spin carriers respectively. The transport spin polarization is the most realistic and relevant parameter from application point of view. The values of $P$ and $P_C$ are same when the Fermi velocities of both the spin currents are equal [87,88].

Point contact Andreev reflection (PCAR) spectroscopy is the most widely used technique to measure $P_C$ in HMFs. In the measurement of $P_C$ by PCAR, one measures the conductance curves across the ferromagnetic (FM)/superconductor (SC) contact and the spin polarization for a ballistic contact can be expressed by equation 10. The nature of the conductance curves for a normal ferromagnet and a HMF is different due to the spin polarization in the latter. In the rest of the discussion, the measured current spin polarization is denoted as $P$ instead of $P_C$. Since this is not a conventional experimental tool, a few details regarding its working are described below.

*PCAR spectroscopy*: This spectroscopy relies on the non-linearity in the current-voltage (I-V) characteristics of a ballistic contact (when contact diameter is less than the electronic mean free (*l*) path, it defined as ballistic contact) with a normal metal (N) on one side and a superconductor (S) on other side. PCAR spectroscopy is somewhat similar to the tunneling spectroscopy, though there are some important differences such as TS is spin independent in the absence of magnetic field while Andreev reflection is spin dependent. Electrical transport through a N/S point contact can be divided into different regimes based on the ratio of the contact diameter (*a*) and the electron mean free path (*l*) [89]. A brief discussion for different regimes is given below.

*The ballistic regime (when a < elastic ($l_e$) or inelastic ($l_i$) scattering)*: In this regime, electrons do not undergo any scattering while passing through the point contact. The conductance is represented by the sum of individual quantized conduction channels. To create a scattering-free region where the kinetic energy of the electron can be increased in a controllable fashion with the help of a voltage is the role of a ballistic contact in point contact spectroscopy. The fundamental principle of PCAR spectroscopy relies on the concept that the kinetic energy of the electron is varied while injecting it from normal metal to a superconductor through a ballistic N/S interface and conductance curves is measured at N/S interface for different kinetic energies. The extreme limit of a ballistic point contact is the quantum point contact. The contact diameter is of the order of (~ 1 – 2 Å) de-Broglie wavelength of the electron in quantum point contact. However, it is very difficult to achieve it in conventional point contact measurements.

*Diffusive regime ($l_{in} > a > l_e$)*: In this regime, there is no inelastic scattering, but electron can undergoe elastic scattering in the point contact region. A diffusive transport comes into play in the spin polarization in itinerant ferromagnets in PCAR.

To make a ballistic point contact which is essential for PCAR, the tip has to be very fine. Different methods are used to make the fine tips depending on the material of the tip. Superconducting metals like niobium (Nb), gold (Au), tungsten (W) etc. can be prepared using electrochemical etching. The starting diameter of the wire is usually 100 – 300 microns. The tip diameter after the processing is reduced to a few tens of nanometers (Fig. 7). However, the final contact diameter is controlled by applying a suitable pressure on the tip. Thus, the pressure on the contact will determine whether it is in ballistic regime or not. Another requirement for obtaining a good ballistic contact is the quality of the sample. The smooth sample surface makes it easier to obtain a ballistic point contact and hence it is often required that sample (polycrystalline bulk/foils/single crystals) surface is cleaned properly and polished to give an almost mirror finish prior to mounting.

The sharp Nb tips prepared using electrochemical polishing is generally used to make superconducting point contacts. Tip is attached to a drive shaft, which is capable of moving up and down, with the help of micrometer mechanism capable of moving the point contact linearly. The transport measurements are made with the conventional four probe arrangement while the point contact and sample are kept at constant temperature (at 4.2 K). The dI/dV data (differential conductance curves) are obtained using the ac lock in techniques at a frequency of 2 kHz. The measured data is analyzed using Blonder, Tinkham, and Klapwijk (BTK) model [88]; a multiple parameter least squares fitting is carried out to deduce spin polarization using dimensionless interfacial scattering parameter (*Z*), superconducting band gap (*Δ*) and *P* as variables The shape of the differential conductance curves depends on the value of *Z*, for low *Z* values curves become flat near *Δ*, while for high *Z* values a peak appear close to *Δ*. The obtained *Δ* values corresponding to the best fit is found to be lower than the bulk superconducting band gap of Nb (1.5 meV), which is attributed to the multiple contacts at the interface [91]. The intrinsic value of current spin polarization is estimated by recording the conductance curve at Z = 0

In the following section, a review of the spin polarization studies reported with the help of PCAR technique on EQHAs is presented. As can be seen, only very few reports are available on the spin polarization measurements, which is an important aspect to be considered for the further progress of this topic.

### A. CoFeMnZ (Z = Si and Ge) alloys

Spin polarization of these alloys were measured using the PCAR technique [90] by Bainsla *et* al. [7,8]. Z = 0.13 was the minimum value that was achieved in the measurement of the differential conductance curves for SGS materials namely CoFeMnSi and CoFeMnGe [7,8]. The estimated current spin polarization value for CoFeMnSi by extrapolating the *P vs. Z* data to *Z* = 0, is shown in the Fig. 8. The current spin polarization value of 0.64 was deduced from the *P vs. Z* curve, which was found to be comparable to the *P* value obtained for some high spin polarization materials using the PCAR technique [9,12-15]. Thus, *P* was found to robust against structural disorder (as discussed above) in the case of CoFeMnSi, as expected for SGS materials [25,26].

The current spin polarization value of 0.70 ± 0.01 was reported from the *P vs. Z* plot for CoFeMnGe by [7], as shown in Fig. 8. The high value of spin polarization obtained for this bulk alloy is found to be comparable to that obtained for bulk $Co_2Fe(Ga_{0.5}Ge_{0.5})$ (P = 0.69 ± 0.02) [15], which has been proved to be one of the best FM Heusler alloys for CPP-GMR devices [18,19]. The work by Bainsla *et* al. on CoFeMnGe[7] appears to be the first spin polarization measurement in any EQHA.

### B. CoFeCrAl alloy

Bainsla *et* al. [9] measured the current spin polarization of CoFeCrAl using PCAR technique (Fig. 8). The estimated *P* = 0.67 ± 0.02 in this alloy is found to be comparable to *that* value reported for CoFeMnGe. The obtained value of *P* for CoFeCrAl is found to be higher than that of many ternary or quaternary Heusler alloys [13,14,92].

In the next section, a summary of various electronic structure calculations reported in different EQHAs is presented.

## VI. ELECTRONIC STRUCTURE CALCULATIONS

Apart from the volume of experimental studies on these alloys, a reasonable amount of work has been reported on their electronic structure by means of theoretical calculations using various tools. Some of the important findings in this regard are presented below.

### A. CoFeMnZ (Z = Al, Si, Ga and Ge) alloys

*Ab initio* electronic structure calculations were performed for CoFeMnZ (Z = Al, Si, Ga, Ge) alloys by Alijani et al. [45]. As mentioned above, depending on the occupancy of the different sites, three different structural configurations are possible for a Y-type structure [45]. The structural optimization using *ab-initio* calculations showed that type I structure with Co(4$d$)Fe($c$)Mn(4$b$)Z(4$a$) exhibit the lowest energy. The total energy for a type I structure was found to be ~270 meV lower than the type II structure. They also compared the optimized lattice parameter ($a_{th}$) values with the experimentally obtained values ($a_{exp.}$) and found that $a_{th}$ values are ≈1% lower than the $a_{exp.}$, as shown in the table 2. The discrepancy between calculated and experimental results could be attributed to the temperature effect (XRD measurement at room temperature), to disorder and/or other structural effects. Thus, Alijani et al. [45] used optimized lattice parameter values for electronic structure and magnetic properties calculations.

The electronic structure calculations showed a typical fully spin polarized half-metallic ferromagnetic band structure for all the alloys. A typical Heusler alloy hybridization gap was observed between the low lying $s$-bands and remaining high lying valence bands (which mainly $p$ and $d$ states) for CoFeMnZ (Z = Al, Ga, Si and Ge) alloys. The small value of sp gap was found for CoFeMnAl alloy which points towards the weak hybridization and thus indicates structural instability. The predicted weak hybridization may be the reason for the experimentally observed antisite disorder in the case of CoFeMnAl (as discussed in the structural properties). Özdogan *et al.* [31] also predicted HMF behavior for CoFeMnZ (Z = Al and Si) using *ab-initio* electronic structure calculations.

Bainsla *et al.* [8] further investigated the electronic structure of the CoFeMnSi alloy using the spin-polarized density functional theory (DFT) within Vienna *ab initio* simulation package (VASP) [93] with a projected augmented-wave basis [94] using GGA exchange correlation functional. In order to make one-to-one comparison between theory and experiment, they used experimental lattice parameter for their calculations. The calculated spin-polarized band structure and density of states (DoS) for an energetically most favorable state are given in Fig. 9. It can be noticed that DoS exhibits a band gap ~0.62 eV for one spin channel, while Fermi level falls within negligible gap in the other spin channel. The unique band structure obtained by Bainsla *et* al. [8] suggests that CoFeMnSi to be a close SGS. CoFeMnSi was predicted to be SGS from another *ab-initio* calculation by Xu *et* al. [30]. The predicted SGS behavior also supports the experimental claims (SGS behavior) made by Bainsla et al. [8]. Therefore, it is understood that to resolve the discrepancy obtained in the findings for CoFeMnSi regarding HMF *vs.* SGS bevaviour, very careful measurements and *ab-initio* calculations are necessary.

Since it has been observed that the thin films based on the Heusler alloys lose their predicted high spin polarization owing to the structural disorder [37-40], Feng *et* al. [95] studied the effect of structural disorder on the electronic and magnetic properties of CoFeMnSi using the *ab-initio* calculations. The effect of various types of antisite and swap disorders on the electronic and magnetic properties of CoFeMnSi alloy was studied comprehensively. The reduction in spin down energy gap of disordered structures was found, and their DoS moves entirely towards the lower zone. Feng *et* al. [95] also found that the HMF nature is maintained under different antisite and/or swap disorders in CoFeMnSi, except for Co(Mn)-, Co(Si)-antisite and Co-Mn, Co-Si swap disorder. Therefore, Feng *et* al. [95] predicted that HMF nature of CoFeMnSi is robust against structural disorder, which supports the experimental observations by Bainsla *et* al. [8].

**B. CoFeCrZ (Z = Al, Ga, Si and Ge) alloys**

Electronic structure of the CoFeCrZ (Z = Al, Ga, Si and Ge) alloys were first studied by Gao et al. [96] using the first-principles full-potential linearized augmented plane-wave (FPLAPW) method with the WIEN2k package [97]. CoFeCrGa and CoFeCrGe are found to have nearly HMF band structure, while CoFeCrAl and CoFeCrSi are reported to exhibit excellent HMF behavior with large half-metallic gaps of 0.16 and 0.28 eV respectively, as per the report of Gao *et* al. [96]. The HMF nature in CoFeCrAl and CoFeCrSi found to be robust against the lattice compression. Gao *et* al. [96], also predicted that the HMF nature of CoFeCrAl and CoFeCrGa is lost, while it is preserved in CoFeCrSi and CoFeCrGe while Coulomb interactions are considered. Xu *et* al. [30], also studied the CoFeCrAl alloy using *ab-initio* calculations and referred it a SGS material. Özdogan *et* al. [31] also predicted CoFeCrAl as having SGS nature, while HMF behavior was predicted for CoFeCrSi.

As discussed above, the contradictory reports on the physical nature (SGS vs. HMF) of CoFeCrAl prompted Bainsla *et* al. [9] to do more careful first principles band structure calculations with spin-polarized DFT using VASP [93,94]. They calculated the spin resolved band structure and DoS of CoFeCrAl for the most stable structural configuration. The calculations were performed with both the experimental lattice parameter ($a_{\text{exp.}}$ = 5.756 Å) and relaxed lattice parameter ($a_{\text{relax}}$ = 5.690 Å), with HMF nature in both cases. The value of the half-metallic band gap was found to be slightly higher (0.37 eV) in the case of $a_{\text{relax}}$ as compared to experimental value (0.31 eV). The band structure calculations exhibit an open band gap (~0.31 eV) for the minority-spin states across the Fermi level ($E_F$), while some valence bands are found to cross the $E_F$ in the majority-spin band and give rise to finite DoS at $E_F$. Therefore, the electronic structure calculations by Bainsla *et* al. [9] suggest HMF nature for CoFeCrAl, which was confirmed by experimental results discussed earlier (electrical resistivity and current spin polarization).

Bainsla *et* al. [10] further investigated the electronic structure of CoFeCrGa alloy. The calculations with $a_{\text{exp.}}$ and most stable structural configuration predicted a closed band gap character for majority-spin state and open band character for minority-spin state near the $E_F$, which suggest that CoFeCrGa behaves like SGS. The value of minority-spin band gap was found to be small (~0.07 eV). They studied the effect of lattice contraction on the electronic and magnetic properties of the CoFeCrGa using *ab-initio* calculations. The predicted DoS versus lattice parameter plots are shown in the Fig. 10. As can be seen, the behavior of CoFeCrGa changes from SGS to HMF with decrease in lattice parameter. Note that the value of half-metallic band gap in minority spin state at $E_F$ was found to increase with decrease in lattice parameter.

Electronic structure of CoFeCrGe was further studied by Enamullah *et* al. [52]. Fully spin-polarized band structure was obtained with finite DoS at $E_F$ for majority-spin states and half-metallic gap in the minority-spin states for CoFeCrGe. The theoretically predicted HMF nature and ordered crystal structure (Y-type) obtained by them make CoFeCrGe a promising candidate for the spintronic applications.

## C. CoMnCrAl, CoFeTiAl and CoMnVAl alloys

Enamullah et al. [52] also calculated the electronic structure of CoMnCrAl using a spin polarized DFT implemented within VASP [93] with a projected augmented-wave basis [94]. HMF band structure and DoS were obtained with experimental lattice parameter and most stable structural configuration. As discussed in the structural properties section, a signature of the $L2_1$ disorder was seen in the room temperature powder XRD patterns of CoMnCrAl by [52]. The electronic structure of HMF materials is generally known to be extremely sensitive to any disorder [37-40]. Therefore, the effect of structural disorder on the stability and electronic structure of CoMnCrAl was studied by Enamullah et al. [52]. HMF nature is found to be preserved under small amounts of structural disorder, while it changes to metallic antiferromagnetic state beyond a certain excess Al in the alloy. Özdogan et al. [31] also predicted HMF nature for CoMnCrAl alloy.

The electronic structure of CoFeTiAl and CoMnVAl was first studied by Basit et al. [51]. Studies on CoMnVAl were carried out in more detail due to its better structural stability (using experimental analysis) and possibility of fully compensated half-metallic ferrimagnetism (FCHMF). The *ab-initio* calculations of the electronic structure for CoMnVAl predicted a non spin-polarized band structure, and this was obtained independent of the starting parameters and methods used (ferromagnetic and/or different type of ferrimagnetic order). Therefore, Basit et al. [51], on the basis of their *ab-initio* calculations, completely excluded the possibility of FCHMF ground state for CoMnVAl. Both CoFeTiAl and CoMnVAl were found to have non spin-polarized band structure [51]. Özdogan et al. [31] predicted that both CoFeTiAl and CoMnVAl behave like semiconductors, similar to the other Heusler alloy with 24 valence electrons. *i.e.* FeMnCrAl, FeMnVSi, FeCrVAs, FeMnTiAs, CoCrVSi, CoMnTiSi, and CoCrTiAs.

Lin et al. [98] studied the effect of anti-site disorder on the electronic and magnetic properties of CoFeTiAl by performing the spin-polarized band structure calculations. An interesting behavior with the antisite disorder was seen by Lin et al. [98]; Co-Ti and Fe-Ti antisite disorder can induce a diluted magnetism in CoFeTiAl. Moreover, the Fe-Ti disorder can induce a 100% spin polarization at $E_F$. Thus, diluted magnetism can be induced in CoFeTiAl by antisite disorder instead of adding a magnetic element.

**D. NiCoMnZ (Z = Al, Ga, Ge and Sn) alloys**

Electronic structure calculations for NiCoMnZ (Z = Al, Ge and Sn) alloys were performed using the FLAPW method by Halder *et* al. [54]. Type I structure {where Ni@4*c*(1/4, 1/4, 1/4), Co@4*d*(/4, 3/4, 3/4), Mn@4*b*(1/2, 1/2, 1/2), and Z@4*c*(0, 0, 0)} was found to be more favorable crystal structure for all the alloys using *ab-initio* calculations. The electronic structure calculations with type I crystal structure and optimized lattice parameter value (5.796 Å) predicted a HMF nature for NiCoMnAl alloy, while a normal ferromagnetic nature was predicted for NiCoMnGe and NiCoMnSn alloys. These calculations suggest a spin polarization value of ~56% and ~60% for NiCoMnGe and NiCoMnSn respectively. The theoretically predicted values of magnetic moment; 5.0, 4.93 and 5.14 $\mu_B$ for NiCoMnAl, NiCoMnGe and NiCoMnSn respectively also indicate the HMF nature in NiCoMnAl. Due to the presence of B2 structural disorder in NiCoMnAl (as discussed in structural properties section), *ab-initio* calculations were performed with B2 disorder, and a pseudogap was seen at approximately 0.15 eV below the Fermi level. Alijani *et* al. [55], calculated the electronic structure of NiCoMnGa alloy. The spin-polarized band structure and DoS calculations were performed with the most stable structural configuration and it was found to exhibit a HMF ground state.

**E. NiFeMnGa and CuCoMnGa alloys**

Electronic structure calculations for these alloys were performed by Alijani *et* al. [55]. The optimization of the cubic structure was the starting point for the calculations. Furthermore, the stability of the cubic structure was studied by calculating the elastic constants $c_{ij}$. On the basis of the $c_{ij}$ values, NiFeMnGa was found to be structurally stable, while CuCoMnGa is structurally unstable. The calculated spin-polarized band structure and DoS predicted a HMF nature for NiFeMnGa, while CuCoMnGa is found to be a normal ferrromagnet.

**F. CoFeTiZ (Z = Si, Ge, Sn and Sb) alloys**

Zhang *et* al. [99] calculated the electronic structure of CoFeTiZ (Z = Si, Ge and Sn) alloys by employing the Cambridge Serial Total Energy Package (CASTEP) [100] based on the DFT plane-wave pseudopotential method. It was found that at the equilibrium lattice parameter, total energy of the ferromagnetic state is lower than that of the paramagnetic state for all these alloys. Thus, Zhang *et* al. [99] performed the calculations while considering the ferromagnetic ground state for all alloys. According to them, CoFeTiSi and CoFeTiGe were found to have half-metallic ferrimagnetic band structure, while CoFeTiSn exhibited a quasi half-metallic ferrimagnetic nature. In the case of CoFeTiSn, Fermi level was found to move slightly into the valence band for the minority spin electrons, which means that it possesses a high spin polarization (nearly 100%), thus it is called as quasi half-metallic ferrimagnet. The total magnetic moment value follows the Slater-Pauling behavior [57-59]. With the variation Z = Si→Ge→Sn, the size of the gap in minority spin band decreases and Fermi level was found to move from the bottom of the minority unoccupied states to the top of the minority occupied states. The half-metallic nature of the alloys was checked under lattice compression (upto -6%) and was found to retain in the case of Z = Si and Ge. In the case of Z = Sn, transition from quasi half-metallic to half-metallic was estimated with lattice compression of ≥1%.

Berri *et al.* [101] calculated the structural, magnetic and electronic properties of CoFeTiSb using full-potential linearized augmented plane wave (FPLAPW) scheme within the GGA. The electronic structure calculations predicted a HMF band structure for CoFeTiSb, with half-metallic gap of 0.53 eV. The calculated total magnetic moment (2.0 $\mu_B$) found to agree with Slater-Pauling rule [57-59].

**G. Other equiatomic quaternary Heusler alloys**

The spin-polarized band structure and DoS calculations were performed in some other EQHAs. *i.e.* CoFeMnZ (Z = As and Sb) by Elahmar *et al.* [102], NiFeTiZ (Z= Si, P, Ge and As) by Karimian *et al.* [103], CoFeScZ (Z = P, As and Sb) by Gao *et al.* [104], CoMnTiZ (Z = P, As and Sb) by Khodami *et al.* [105], NiCoCrGa by Gökhan *et al.* [106],VCoHfGa and CrFeHfGa by Yang *et al.* [107], and FeCrMnSb by Mukhtiyar *et al.* [108]. The results of these calculations are summarized in the table 4. Özdogan *et al.* [31], also performed *ab-initio* calculations on various EQHA systems in order to understand their HMF and/or SGS properties. Table 4 summarized the results of theoretical calculations on different EQHA systems.

**VII. SUMMARY AND CONCLUSIONS**

Equiatomic quaternary Heusler alloys with Y-type crystal structure show interesting properties useful for applications in spintronic devices. They have been subjected to a large number of experimental and theoretical investigations in view of their fundamental and applied interests. Possibilities of achieving high Curie temperature and large spin polarization are the most crucial aspects of these alloys in general. They also have the advantage of larger spin diffusion length over pseudo-ternary Heusler alloys such as $X_2Y_{1-a}Y'_aZ$, since the random distribution of Y and Y' in the latter gives additional disorder scattering. The electronic structure and hence the physical properties of these alloys strongly depend on the structural order and distribution of the atoms in the crystal lattice. Tunability of the electronic structure and thereby the magnetic and electrical properties is another advantage of EQHAs. Therefore, these alloys constitute one of the most important material systems that belong to the half-metallic ferromagnetic family. A special class of HMF materials, known as spin gapless semiconductors is also drawing a lot of attention from researchers currently. One can expect very interesting materials from this category as well in the very near future.

Various studies reported so far indicate that achieving a fully ordered crystal structure in EQHAs is almost impossible. It was found that even a small amount of atomic disorder could cause changes in their electronic structure and their magnetic and transport properties. Therefore, the role of disorder in these materials is a crucial factor to be assessed before investigating its application potential. Experimental tools such as powder XRD, Mössbauer spectroscopy, neutron diffraction and EXAFS are being used to identify the crystal structure of EQHAs and to estimate the amount of disorder. Generally, the magnetic moment of these alloys can be estimated using the Slater-Pauling rule. The total magnetic moment is found to have integer values for HMF electronic structure. High value of saturation magnetization and Curie temperature were obtained for some EQHAs. A rough idea about the existence of the HMF nature can be obtained using the electrical resistivity analysis. However, Point contact Andreev reflection measurement appears to be the most reliable, widely used and practical method to estimate the degree of spin polarization, both in bulk and thin films forms. Hall measurements along with PCAR constitute the most important characterization techniques to probe SGS behavior.

The current spin polarization value, $P = 0.70 \pm 0.01$ for CoFeMnGe is found to be highest among the EQHAs reported so far. This value is slightly better than the value ($P = 0.69 \pm 0.02$) obtained for bulk $Co_2FeGa_{0.5}Ge_{0.5}$ alloy, which is heavily used in the spintronic devices. In general, Co-based alloys have high $T_C$ values, eg: CoRuFeSi has the highest $T_C$ among all the EQHAs. In view of their excellent properties, these materials are considered to be promising for various spintronic applications. SGS materials are found to be suitable to replace diluted magnetic semiconductors. Though at present SGS behavior is experimentally confirmed only for CoFeMnSi and CoFeCrGa, the results regarding the robustness of their properties against structural disorder are very encouraging. Therefore, it is clear that EQHAs offer many advantageous properties, which can be exploited for many applications in spintronics.

**ACKNOWLEDGEMENTS**

The authors thank Prof. K. Hono (NIMS, Japan), Prof. A. Alam (IIT Bombay), Prof. A. K. Nigam (TIFR, India) for their help. KGS acknowledges the financial support received from ISRO Cell (IITB) through a sponsored project.

**Tables**

TABLE I. Site occupancy, general formula and structure type (according to different databases) for different atomic orders of Heusler alloys. Space groups corresponding to various structures are also given.

| Site occupancy | General Formula | Structure type ICSD | Strukturberichte (SB) database | Space group |
|---|---|---|---|---|
| X, X', Y, Z | XX'YZ | LiMgPdSn | Y | F-43m (# 216) |
| X = X, Y, Z | $X_2YZ$ | $Cu_2MnAl$ | $L2_1$ | Fm-3m (# 225) |
| X = X', Y = Z | $X_2Y_2$ | CsCl | B2 | Pm-3m (# 221) |
| X = X' = Y, Z | $X_3Z$ | $BiF_3$ | $DO_3$ | Fm-3m (# 225) |
| X = X' = Y = Z | $X_4$ | W | A2 | Im-3m (# 229) |

TABLE II. Structure type, crystal structure (ordered/disorder), experimental lattice parameter ($a_{exp.}$), space group and corresponding references for various EQHAs.

| Alloy | Structure type | Crystal structure (ordered/disorder) | $a_{exp.}$ (Å) | Space group | Reference |
|---|---|---|---|---|---|
| CoFeMnAl | | B2 disorder | 5.79 | | [45] |
| CoFeMnSi | | $DO_3$ disorder | 5.66 | | [8], [45] |
| | | | 5.66 | | |
| CoFeMnGa | | to be probed in detail | 5.81 | | [45] |
| CoFeMnGe | | ordered (Y-type) | 5.76 | | [7], [45] |
| CoFeCrAl | | B2 disorder | 5.76 | | [9], [47], [48] |
| CoFeCrGa | | DO$_3$ disorder | 5.79 | | [10] |
| CoFeCrGe | | ordered (Y-type) | 5.77 | | [52] |
| CoFeTiAl | | ordered (Y-type) | 5.85 | | [51] |
| CoMnVAl | | ordered (Y-type) | 5.80 | | [51] |
| CoMnCrAl | | L2$_1$ disorder | 5.76 | | [52] |
| CoRuFeSi | Y | B2 disorder | 5.77 | $F\bar{4}3m$ | [11], [50] |
| CoRuFeGe | | B2 disorder | 5.87 | | [11] |
| NiCoMnAl | | B2 disorder | 5.79 | | [54] |
| NiCoMnGe | | ordered (Y-type) | 5.78 | | [54] |
| NiCoMnSn | | ordered (Y-type) | | | [54] |
| NiCoMnGa | | to be probed in detail | 5.80 | | [55] |
| NiFeMnGa | | | 5.80 | | [55] |

| Alloy | | | 5.85 | | [55] |
|---|---|---|---|---|---|
| CuCoMnGa | | | | | |
| MnNiCuSb | | ordered (Y-type) | 5.92 | | [56] |

TABLE III. List of Curie temperature ($T_C$), saturation magnetization ($M_S$), magnetic moment value using Slater-Pauling rule ($M_{S-P}$) and calculated magnetic moment ($M_C$) with references for some EQHAs.

| Alloy | $M_S$ ($\mu_B$/f.u.) [experimental] | $M_C$ ($\mu_B$/f.u.) [calculated] | $M_{S-P}$ ($\mu_B$/f.u.) | $T_C$ (K) | References |
|---|---|---|---|---|---|
| CoFeMnAl | 3.1 (at 5 K) and 2.7 (at 300 K) | 3.0 | 3.0 | 553 | [45] |
| CoFeMnSi | 4.1 (at 5 K) and 3.7 (at 300 K) 3.7 (at 3 K) | 4.0 4.0 | 4.0 | 623 | [45] [8] [8] |
| CoFeMnGa | 3.2 (at 5 K) and 2.8 (at 300 K) | 3.0 | 3.0 | 567 | [45] |
| CoFeMnGe | 4.2 (at 3 K) and 3.8 (at 300 K) | 4.0 | 4.0 | 711 | [8], [45] |
| CoFeCrAl | 2.0 | 2.0 | 2.0 | 460 | [9] |
| CoFeCrGa | 2.1 | 2.0 | 2.0 | above 400 | [10] |
| CoFeCrGe | 3.0 | 3.0 | 3.0 | 866 | [52] |
| CoFeTiAl | 0 | | 0 | | [51] |
| CoMnVAl | 0 | | 0 | | [51] |

| Composition | | | | | |
|---|---|---|---|---|---|
| CoMnCrAl | 0.9 | 1.0 | 1.0 | 358 | [52] |
| CoRuFeSi | 4.8 | | 5.0 | 867 | [11] |
| CoRuFeGe | 5.0 | | 5.0 | 833 | [11] |
| NiCoMnAl | 4.7 | 5.0 | 5.0 | | [54] |
| NiCoMnGa | 4.5 | 5.1 | 5.0 | 646 | [55] |
| NiCoMnGe | 4.1 | 4.9 | 6.0 | | [54] |
| NiCoMnSn | 4.8 | 5.1 | 6.0 | | [54] |
| NiFeMnGa | 3.4 | 4.0 | 4.0 | 326 | [55] |
| CuCoMnGa | 2.3 | 4.3 | 4.0 | 631 | [55] |
| MnNiCuSb | 3.9 | | | 690 | [56] |

TABLE IV. Results of the electronic structure and magnetic properties calculations for various EQHAs, with composition, theoretical lattice parameter ($a_{th.}$), calculated magnetic moment ($M_C$), and magnetic moment estimated using Slater-Pauling rule ($M_{S-P}$), physical nature and corresponding references.

| Composition | $a_{th.}$ (Å) | $M_C$ (μB/f.u.) | $M_{S-P}$ (μB/f.u.) | Physical nature | References |
|---|---|---|---|---|---|
| CoFeMnAl | 5.692 | 3.0 | 3.0 | HMF | [45] |
| CoFeMnSi | 5.610 | 4.01~4.0 | 4.0 | SGS | [8] |
| CoFeMnGa | 5.708 | 3.0 | 3.0 | HMF | [45] |
| CoFeMnGe | 5.710 | 4.0 | 4.0 | HMF | [7] |
| CoFeMnAs | 5.740 | 4.99~5.0 | 5.0 | HMF | [102] |
| CoFeMnSb | 5.981 | 5.0 | 5.0 | HMF | [102] |
| CoFeCrAl | 5.69 | 1.97 | 2.0 | HMF | [9], [96] |
| CoFeCrSi | 5.63 | 3.0 | 3.0 | HMF | [31], [96] |
| CoFeCrGa | 5.73 | 2.0 | 2.0 | HMF | [10] |
| CoFeCrGe | 5.72 | 2.99~3.0 | 3.0 | HMF | [52], [96] |
| CoMnCrAl | 5.71 | 0.98 | 1.0 | HMF | [31], [52] |
| CoMnVAl | - | 0 | 0 | SC | [31], [51] |
| NiCoMnAl | - | 5.0 | 5.0 | HMF | [54] |
| NiCoMnGa | 5.784 | 5.07 | 5.0 | HMF | [55] |
| NiCoMnGe | 5.796 | 4.93 | 6.0 | FM | [54] |
| NiCoMnSn | - | 5.14 | 6.0 | FM | [54] |
| NiFeMnGa | 5.755 | 4.01~4.0 | 4.0 | HMF | [55] |
| CuCoMnGa | 5.846 | 4.32 | 4.0 | FM | [55] |
| CoFeTiAl | - | 0 | 0 | SC | [31], [51] |
| CoFeTiSi | 5.73 | 1.0 | 1.0 | HMF | [99] |
| CoFeTiGe | 5.81 | 1.0 | 1.0 | HMF | [99] |
| CoFeTiSn | 6.09 | 1.0 | 1.0 | quasi HMF | [99] |
| CoFeTiSb | 6.08 | 2.0 | 2.0 | HMF | [101] |
| NiFeTiSi | 5.84 | 2.0 | 2.0 | HMF | [103] |
| NiFeTiGe | 5.93 | 1.77 | 2.0 | HMF | [103] |
| NiFeTiP | 5.78 | 1.0 | 1.0 | HMF | [103] |
| NiFeTiAs | 5.83 | 1.39 | 1.0 | FM | [103] |
| CoFeScP | 5.789 | 1.0 | 1.0 | HMF | [104] |
| CoFeScAs | 5.942 | 1.01~1.0 | 1.0 | HMF | [104] |
| CoFeScSb | 6.198 | 1.02 | 1.0 | HMF | [104] |
| CoMnTiP | 5.69 | 1.0 | 1.0 | HMF | [105] |
| CoMnTiAs | 5.83 | 1.0 | 1.0 | HMF | [105] |
| CoMnTiSb | 6.07 | 1.0 | 1.0 | HMF | [105] |
| NiCoCrGa | 5.781 | 4.01 | 4.0 | HMF | [106] |
| VCoHfGa | 6.261 | 3.02 | 3.0 | SGS | [107] |
| CrFeHfGa | 6.261 | 3.0 | 3.0 | SGS | [107] |
| FeCrMnSb | 6.067 | 2.0 | 2.0 | HMF | [108] |

Özdogan *et* al. [31], also performed *ab-initio* calculations on various EQHA systems in order to view their HMF and/or SGS properties.

**Figures**

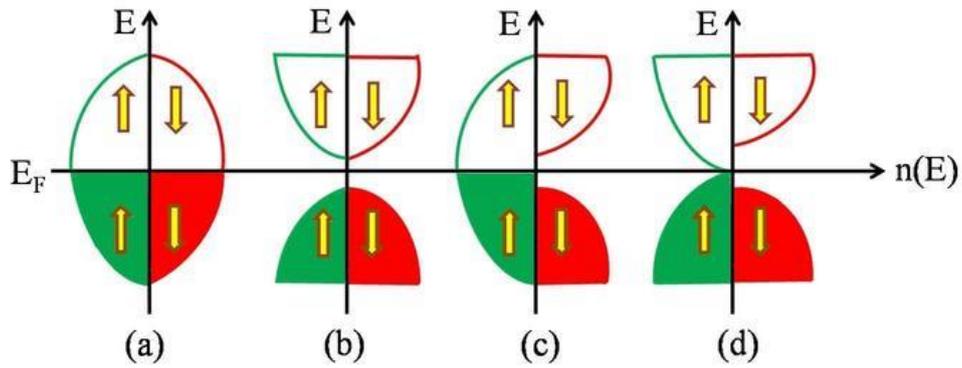

FIG. 1. Schematic of density of states for a typical (a) nonmagnetic metal, (b) semiconductor (c) half-metallic ferromagnet and (d) spin gapless semiconductor. Reprinted with permission from Bainsla *et* al., Phys. Rev. B 91, 104408 (2015). Copyright 2015 American Physical Society.

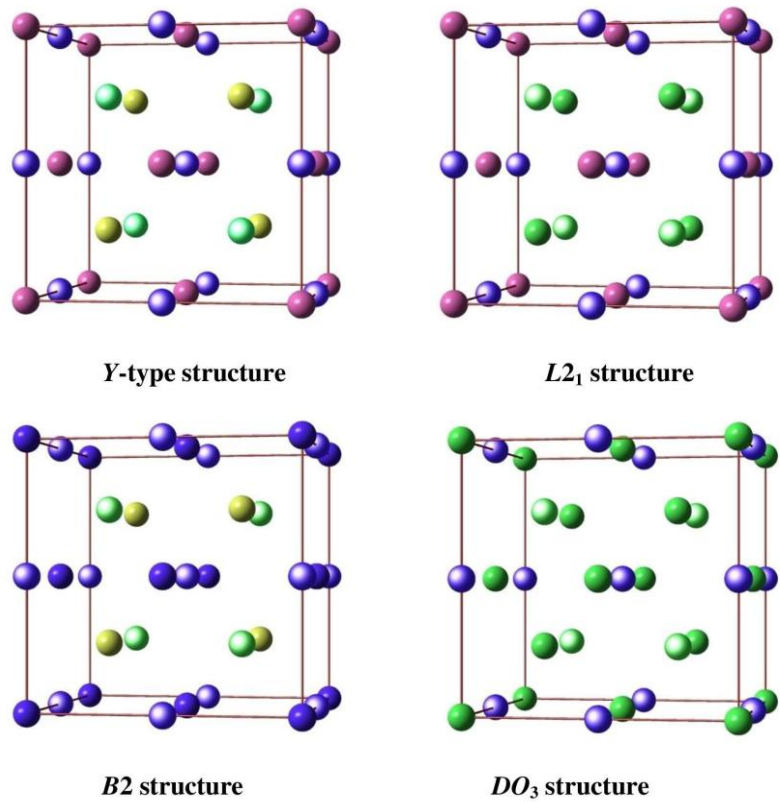

FIG. 2. An overview of the different types of crystal structures in Heusler alloys. Blue, violet, yellowish green and green circles represent the $4a(0, 0, 0)$, $4b(1/2, 1/2, 1/2)$, $4c(1/4, 1/4, 1/4)$ and $4d(3/4, 3/4, 3/4)$. Considering, X@$4c(1/4, 1/4, 1/4)$, X'@$4d(3/4, 3/4, 3/4)$, Y@$4b(1/2, 1/2, 1/2)$ and Z@$4a(0, 0, 0)$ for an EQHA with Y-type structure.

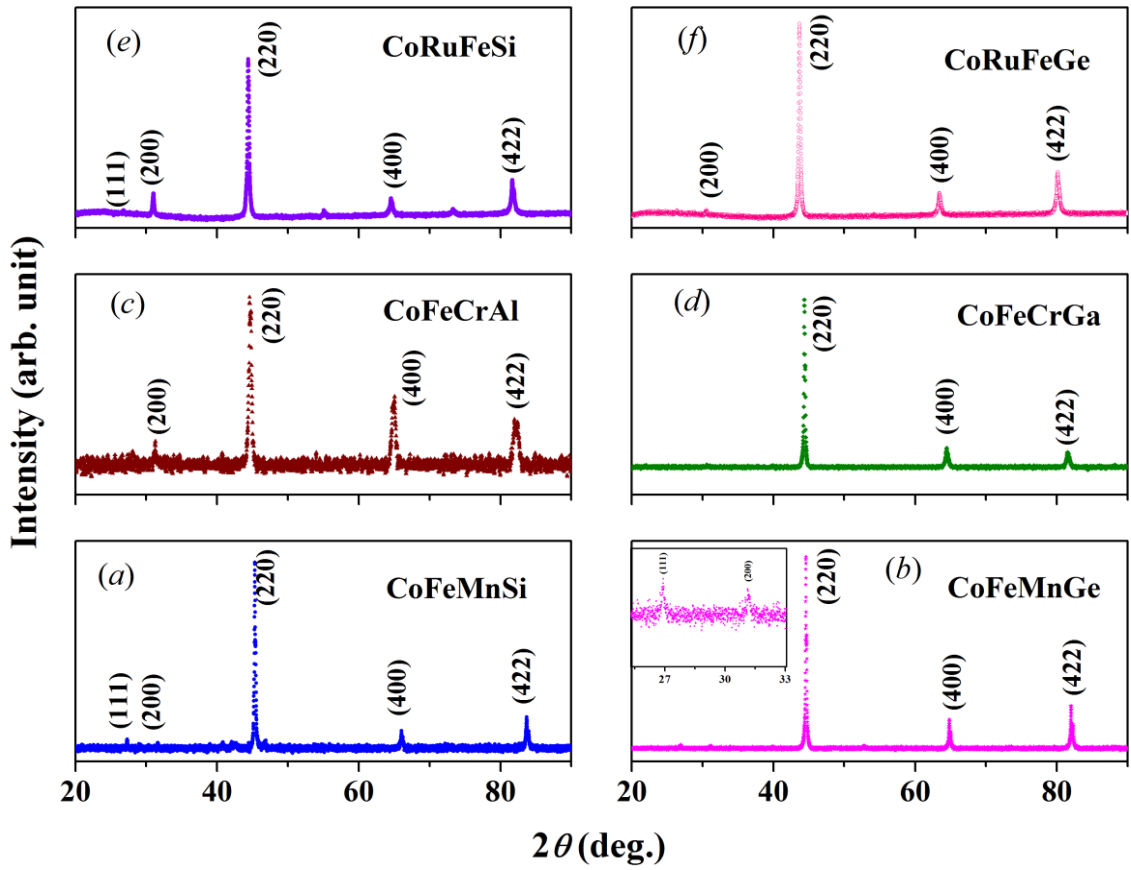

FIG. 3. Room temperature powder x-ray diffraction patterns of some EQHAs. (a) and (b) show the pattern obtained for the CoFeMnX (X = Si and Ge) alloys. The inset in Fig. (b) shows the zoomed version in the region of 26-33 degree. (c), (d), (e) and (f) show the patterns for CoFeCrAl, CoFeCrGa, CoRuFeSi and CoRuFeGe respectively.

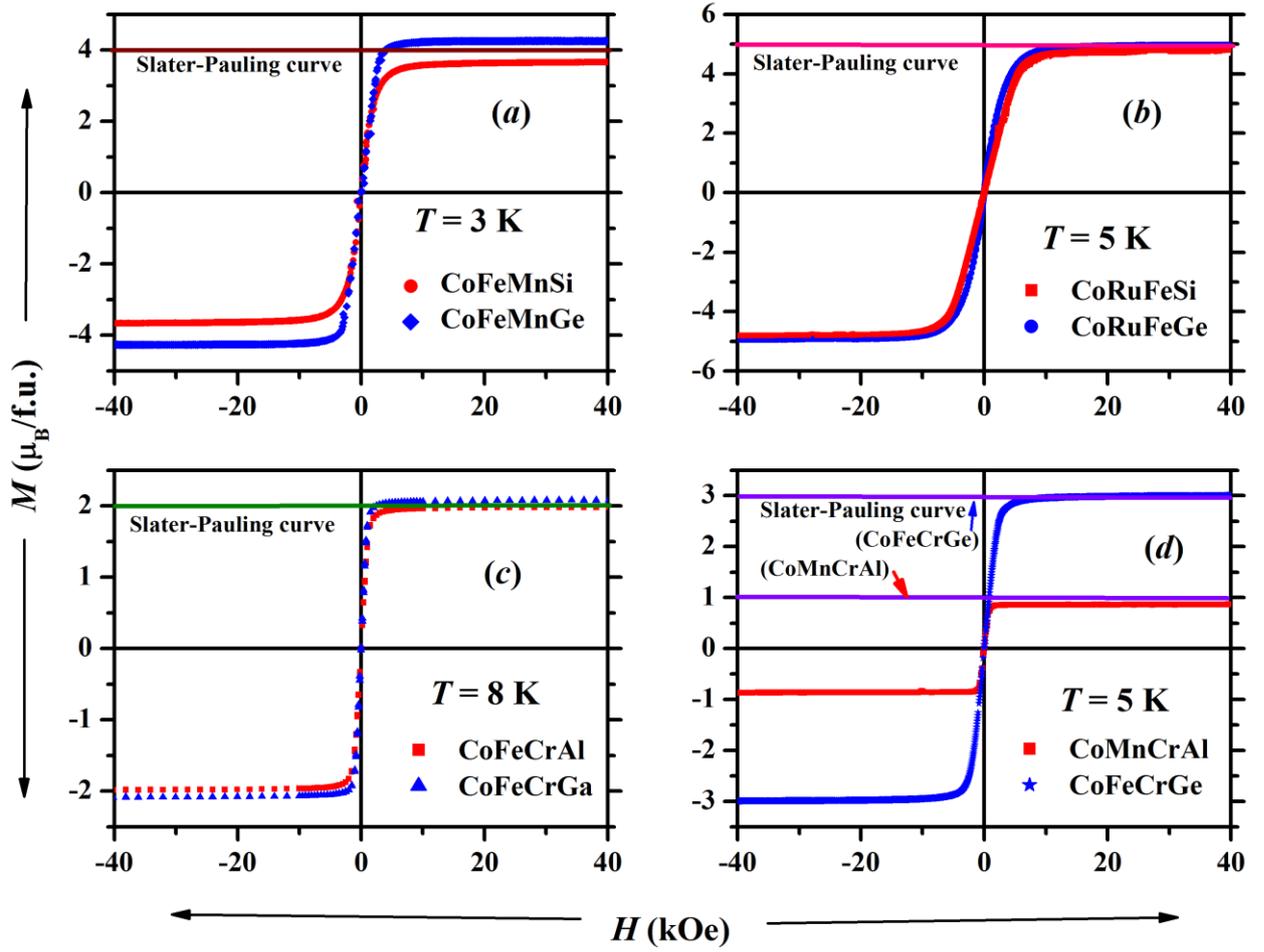

FIG. 4. Isothermal magnetization curves obtained for certain equiatomic quaternary Heusler alloys at low temperatures. Fig. (*a*) and (*b*) represents the *M vs. H* curves for CoFeMnZ (Z = Si and Ge) and CoRuFeZ (Z = Si and Ge) respectively. (*c*) shows the curves for CoFeCrZ (Z = Al and Ga) alloys at 8 K, while (*d*) shows the isothermal curves for CoMnCrAl and CoFeCrGe alloys at a temperature of 5 K.

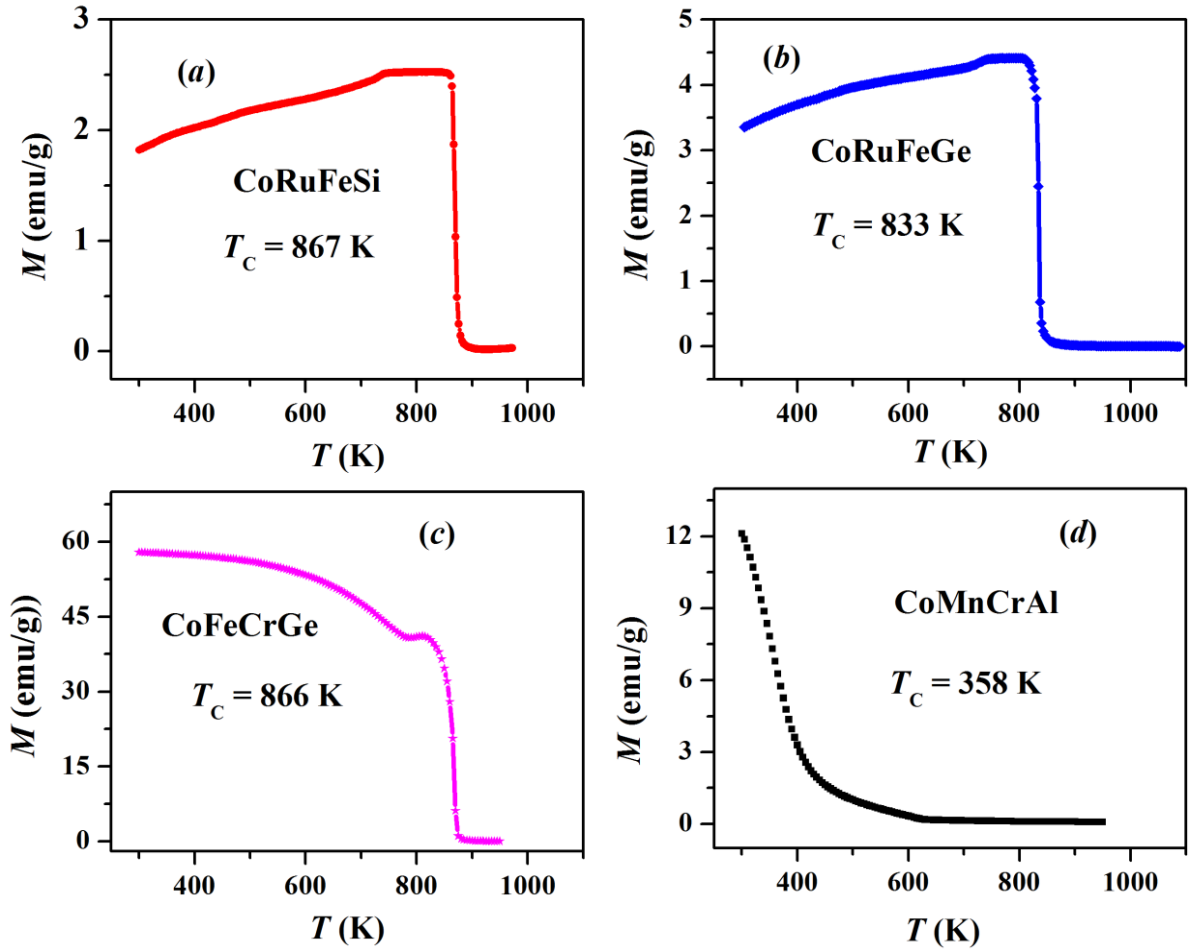

FIG. 5. Thermomagnetic curves obtained for different equiatomic quaternary Heusler alloys in the range of 300 – 1100 K. Fig. (*a*) and (*b*) represents the curves for CoRuFeSi and CoRuFeGe respectively. While, (*c*) and (*d*) show the curves for CoFeCrGe and CoMnCrAl respectively.

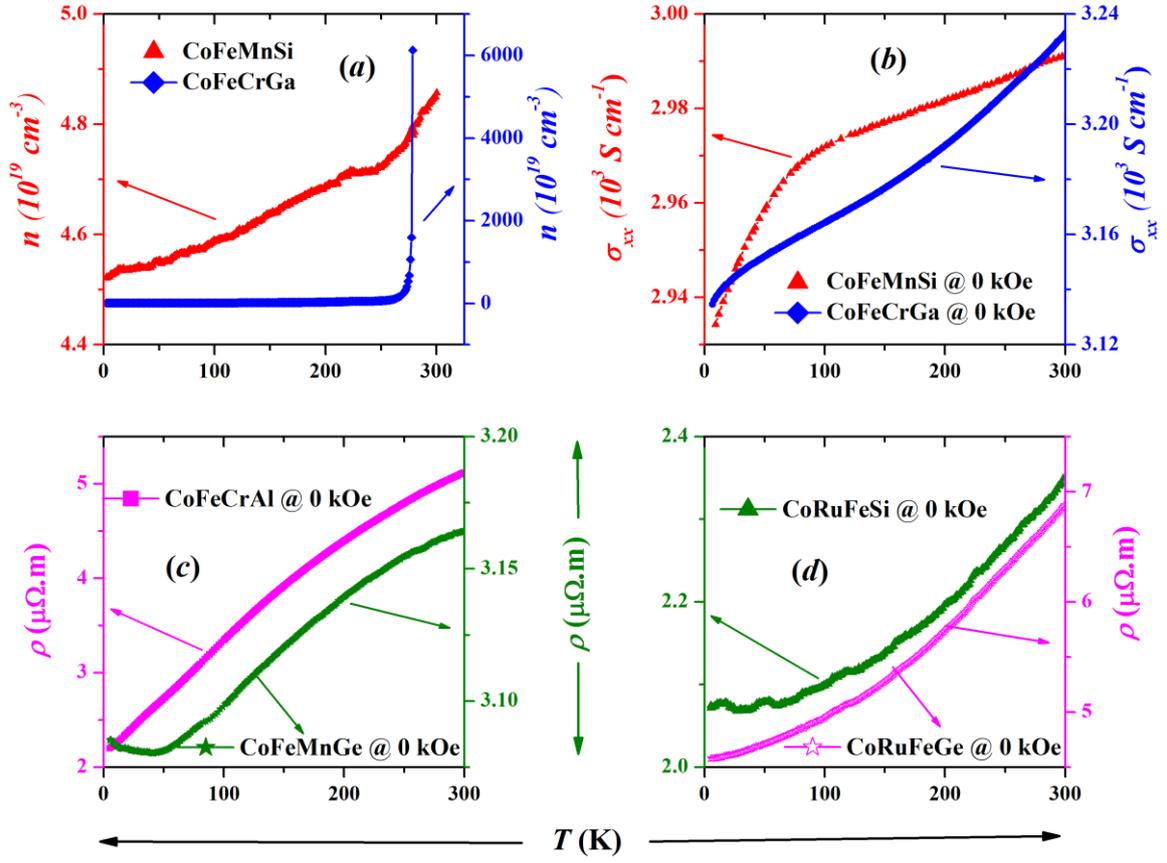

FIG. 6. (*a*) and (*b*) represent the carrier concentration and electrical conductivity curves for the CoFeMnSi and CoFeCrGa alloys respectively, in the temperature range of 5 – 300 K. (c) shows the electrical resistivity curves for the CoFeCrAl and CoFeMnGe alloys, while (*d*) shows the electrical resistivity curves for the CoRuFeZ (Z = Si and Ge) alloy.

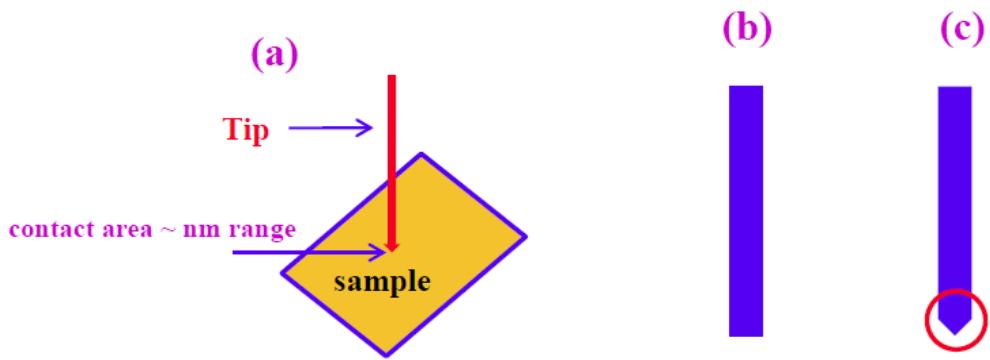

FIG. 7. (a) Sample-tip configuration in PCAR (b) a cartoon showing the original superconducting wire. (c) superconducting wire after electrochemical etching.

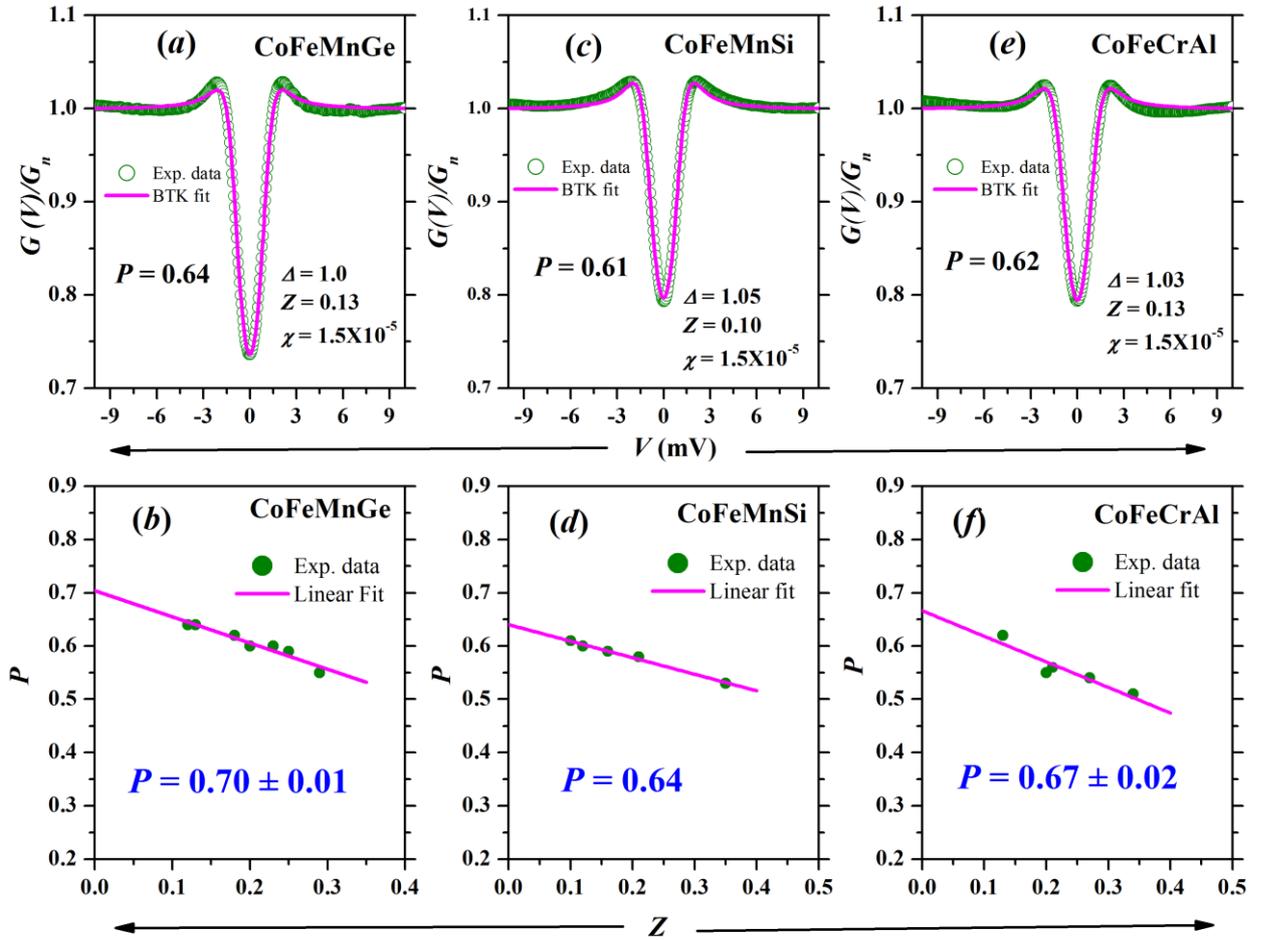

FIG. 8. (*a*), (c) and (*e*) show the normalized differential conductance curves corresponding to the lowest achievable Z values at 4.2 K for CoFeMnGe, CoFeMnSi and CoFeCrAl respectively. (*b*), (*d*) and (*f*) represent the *P vs. Z* curves for CoFeMnGe, CoFeMnSi and CoFeCrAl respectively.

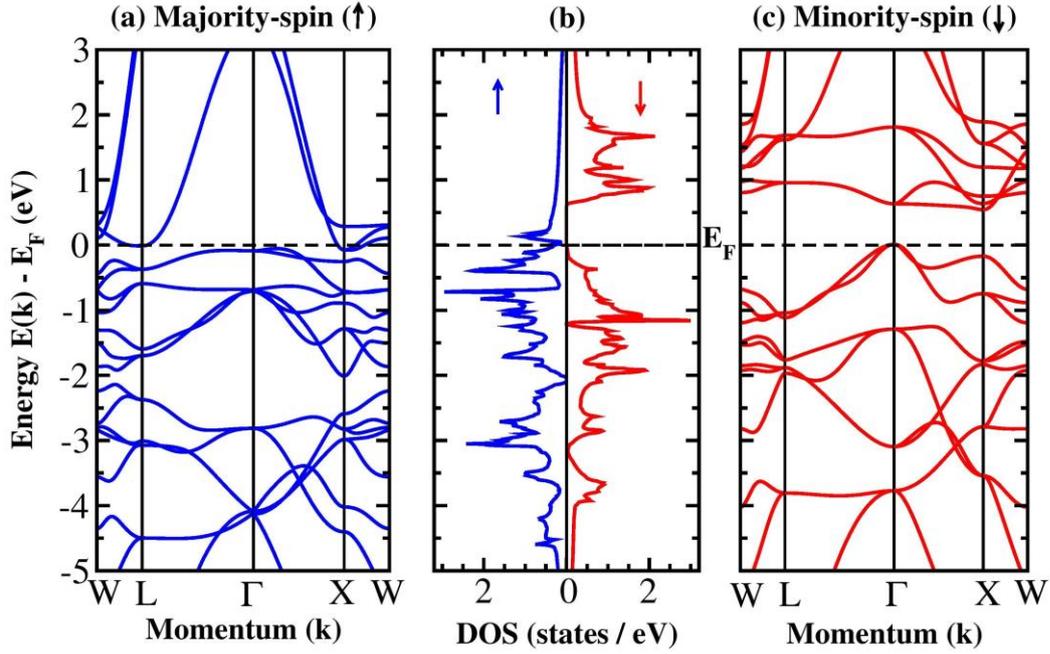

FIG. 9. Band structure and density of states for CoFeMnSi. (a) majority-spin band, (b) density of states, (c) minority-spin states. Reprinted with permission from Bainsla *et* al., Phys. Rev. B 91, 104408 (2015). Copyright 2015 American Physical Society.

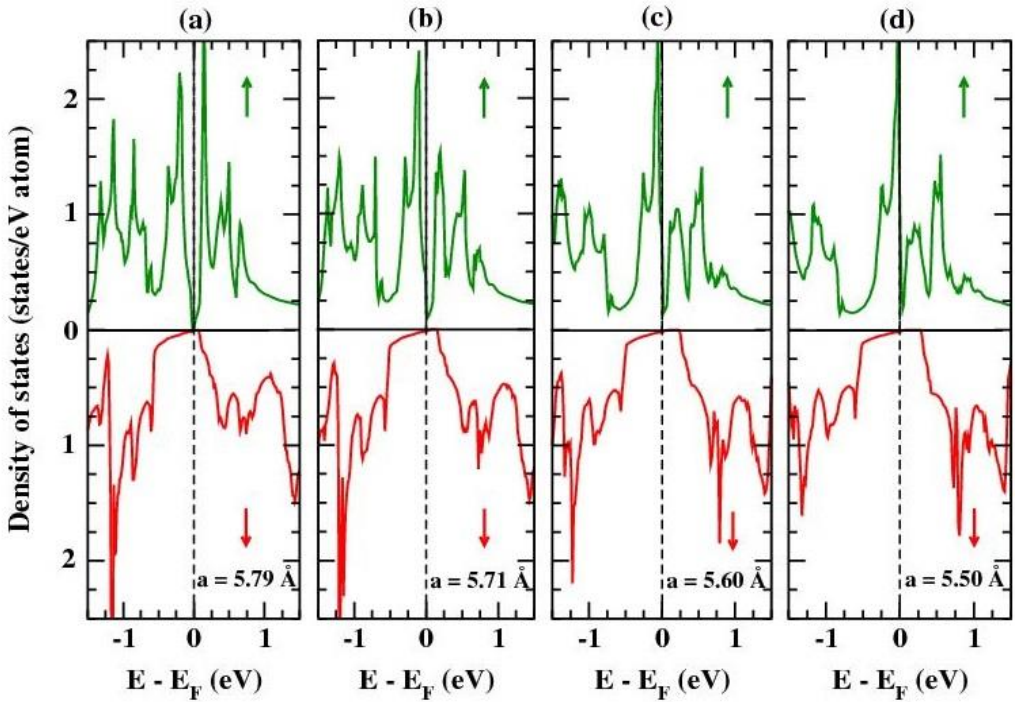

FIG. 10. DoS *vs.* $a_{exp.}$ for CoFeCrGa alloy. (a) 5.79 Å ($a_{exp.}$), (b) 5.71 Å, (c) 5.60 Å, and (d) 5.50 Å. Reprinted with permission from Bainsla *et* al., Phys. Rev. B 92, 045201 (2015). Copyright 2015 American Physical Society.